\documentclass[aps,prb,reprint,twocolumn,longbibliography]{revtex4-2}
\pdfoutput=1

\usepackage{graphicx}
\usepackage[space]{grffile}
\usepackage{amsmath,amsfonts,amssymb,mathtools}
\usepackage{color}
\usepackage{subfigure}
\usepackage{dsfont}
\usepackage{hyperref}
\usepackage{booktabs}
\usepackage{url}
\usepackage{leftidx}
\usepackage{textcomp}
\usepackage{gensymb}
\usepackage{threeparttable}
\usepackage{rotating}
\usepackage{esvect}
\usepackage{booktabs}
\usepackage{lmodern}
\usepackage[noend]{algpseudocode}
\usepackage{adjustbox}
\usepackage{multirow}
\usepackage{svg}
\usepackage[normalem]{ulem}
\usepackage{braket}
\usepackage{bm}

\newcommand{\ve}[1]{{\bm{#1}}}

\graphicspath{{../Inkscape/},{../Mathematica/},{./Figures/},{./images/}}

\begin{document}

\title{Interplay of superconductivity and charge-density-wave order in kagome materials}

\author{Sofie Castro Holb{\ae}k}
\affiliation{Department of Physics, University of Zurich, Winterthurerstrasse 190, 8057 Zurich, Switzerland}

\author{Mark H. Fischer}
\affiliation{Department of Physics, University of Zurich, Winterthurerstrasse 190, 8057 Zurich, Switzerland}

\date{\today}

\begin{abstract}
In the \textit{A}V$_{3}$Sb$_{5}$ (\textit{A}~$=$~K,~Rb,~Cs) kagome materials, superconductivity coexists with a charge density wave (CDW), constituting a new platform to study the interplay of these two orders. Despite extensive research, the symmetry of the superconducting order parameter remains disputed, with experiments seemingly supporting different conclusions. As key aspects of the physics might lie in the intertwining of electronic orders, a better understanding of the impact of the CDW on superconductivity is crucial. In this work, we develop a phenomenological framework to study the interplay of superconductivity and CDW order. In particular, we derive a Ginzburg-Landau free energy for both superconducting and CDW order parameters. Given the unclear nature of the superconducting state, we discuss general pairing symmetries with a focus on $s$-wave, $d$-wave, and pair-density-wave order parameters. Motivated by experiments, we consider the additional breaking of time-reversal or point-group symmetries of the CDW and determine in detail the consequences for the superconducting state. Our results show how the superconducting state mimics the broken symmetries of the CDW and can guide future microscopic calculations, as well as the experimental identification of the superconducting state in the \textit{A}V$_{3}$Sb$_{5}$ compounds.
\end{abstract}

\maketitle

\section{Introduction}
In many well-known materials, superconductivity develops on the background of a charge-density-wave (CDW) order. Notable examples include various cuprate superconductors, such as YBa$_{2}$Cu$_{3}$O$_{7-x}$~\cite{chang:2012}, and transition metal dichalcogenides including 2$H$-NbSe$_{2}$~\cite{cho:2018}, 1$T$-TiSe$_{2}$~\cite{kusmartseva:2009}, and 2$H$-TaS$_{2}$~\cite{freitas:2016}. Especially in the context of the cuprate superconductors, it is argued that studying the intertwining of various orders is crucial to understanding the intricate phase diagram~\cite{fradkin:2015}. Finally, in many of the above systems, the breaking of translational symmetry by the charge density wave was reported to modulate the superconducting order~\cite{hamidian:2016,du:2020,liu:2021,gu:2023,liu:2023,zhao:2023}, a clear indication of interaction between the orders.

A novel system was added to this list with the recently discovered \textit{A}V$_{3}$Sb$_{5}$ (\textit{A}VS) (with \textit{A}~$=$~K,~Rb,~Cs) family of kagome metals~\cite{ortiz:2019,ortiz:2020}, again featuring superconductivity coexisting with a charge density wave. All three \textit{A}VS members enter a charge-ordered state at a transition temperature of $\sim 80$--$100$~K---the specific value depending on the alkali atom~\cite{uykur:2021,uykur:2022,wenzel:2022}. For all three compounds, the observed in-plane charge redistribution leads to a $2\times 2$ increase of the in-plane unit cell~\cite{jiang:2021} corresponding to a modulation with wave vector $\ve{q}=\ve{M}_{i}$, the three M-points of the hexagonal Brillouin zone. The out-of-plane periodicity, in contrast, differs between the three compounds~\cite{kautzsch:2023,frassineti:2023,kang:2023,stahl:2022,hu:2022,xiao:2023} featuring either a $2\times2\times2$ or a $2\times2\times4$ modulation in three dimensions.

Importantly, unlike most other examples, where the CDW is expected to be of a rather conventional nature, several experiments in \textit{A}VS indicate that multiple symmetries beyond translational symmetry are spontaneously broken in the CDW phase. These include the breaking of three-fold rotational symmetry, (switchable) structural chirality in the form of inequivalent CDW peaks~\cite{jiang:2021,shumiya:2021,wang:2021,xing:2024}, and signs of broken time-reversal symmetry in the absence of local moments~\cite{xu:2022,mielke:2022,khasanov:2022,guguchia:2023}. The latter has been interpreted as evidence of orbital currents, or flux order, in addition to the more conventional bond order. Note, however, that other experiments challenge the intrinsic nature of some of these measurements~\cite{li:2022,li:2022b,saykin:2023,wang:2024}, and, in particular for the switchable chirality, stress the importance of extrinsic effects, such as applied magnetic fields and strain~\cite{guo:2024,xing:2024}. Given its rich CDW physics, the \textit{A}VS family presents an intriguing new platform for studying the interplay of various charge orders with superconductivity.

The kagome metals of the \textit{A}VS family finally become superconducting at a critical temperature $T_{\rm c}\sim 1$--$3$~K \cite{ortiz:2021,ortiz:2020,yin:2021}.
Despite its more complex phase diagram, most experiments probing the superconducting state have focused on CsV$_{3}$Sb$_{5}$, which features the highest critical temperature among the pristine members with $T_{\rm c} = 2.5$~K. Several experiments describe the superconducting gap as nodeless, anisotropic, and with singlet pairing~\cite{duan:2021,roppongi:2023,zhao:2023,zhong:2023,zhang:2023}, and given the multiple bands at the Fermi level, two-gap models have been proposed to accurately describe experimental observations~\cite{hossain:2024,gupta:2022,shan:2022,xu:2021}. Furthermore, recent angle-resolved photoemission spectroscopy (ARPES) measurements on CsV$_{3}$Sb$_{5}$ suggest an isotropic gap on the Sb-derived Fermi surface and a highly anisotropic gap on the V-derived hexagonal Fermi surface, where the CDW plays a more significant role~\cite{mine:2024}. Pressure studies that suppress the CDW have shown a reduction in anisotropy and a transition to an isotropic gap~\cite{guguchia:2023}, highlighting the possibly significant role of the CDW in the observed superconductivity at ambient pressure.
Interestingly, also in this family of materials, a modulation of superconductivity was recently reported~\cite{deng:2024,yan:2024}.

The discovery of CDW order and superconductivity in the \textit{A}VS kagome metals has stimulated a significant effort to determine the microscopic structure and origin of the ordered phases. For this purpose, one often studies charge density waves or superconductivity on the kagome lattice independently. Focusing on unconventional pairing interactions, several theoretical studies have suggested a close competition of various channels with an extended $s$-wave or a chiral $d$-wave state often the leading instability~\cite{yu:2012a,wang:2013c,kiesel:2013a,wu:2021,wen:2022,tazai:2022,romer:2022}. In contrast, an (orbital-selective) phonon-mediated pairing attraction was suggested in Ref.~\onlinecite{ritz:2023}. However, the interplay of orders can lead to new phenomena that cannot be understood from studying each order by itself~\cite{fradkin:2015}. Therefore, examining the mutual influence of individual orders can be crucial for understanding the phase diagram, and additionally provide insights into the mechanism driving each phenomenon.

Previous studies including both CDW order and superconductivity have mainly focused on simplified microscopic models~\cite{jiang:2023, jiang:2024, lin:2024} or within Ginzburg-Landau theory restricted to specific charge-order and pairing symmetries~\cite{yang:2023}. Stipulating on-site attraction and thus, conventional $s$-wave pairing, a time-reversal-symmetry-breaking flux phase can induce anisotropy in the excitation gap~\cite{jiang:2023, jiang:2024} and even lead to a topologically non-trivial state~\cite{lin:2024}, while the system with a bond-ordered CDW retains an isotropic gap. Considering only a flux phase and superconductivity within a patch model valid close to the van Hove singularity, the interaction of $s$- or $d$-wave pairing was, further, investigated within Ginzburg-Landau theory~\cite{yang:2023}. Finally, a recent study explored the possibility of a pair-density-wave instability and the resulting orders interacting with a charge density wave within a self-consistent mean-field theory~\cite{yao:2024tmp}. Yet, no full classification of superconducting orders and their interplay with various forms of charge order has been presented up to date.

Here, we wish to comprehensively study the interplay of charge-density-wave and superconducting orders on the kagome lattice, including their potential for new states. While indications of additional order with different periodicity were reported in surface-sensitive probes~\cite{zhao:2021,li:2023,chen:2021b}, we focus on a commensurate (in-plane) CDW leading to a $2\times 2$ enlarged unit cell consistent with X-ray studies in all three compounds~\cite{li:2021,ortiz:2021b,li:2022c} and study its interplay with homogeneous, in other words total momentum $\ve{q}=\ve{0}$, superconductivity and pair-density-wave order of the same wave vectors as the CDW, $\ve{q}=\ve{M}_{i}$.
In the spirit of the framework recently introduced in Ref.~\onlinecite{venderbos:2016}, we derive the Ginzburg-Landau (GL) free energy of possible CDWs and superconducting orders and explore the consequences of the dominant coupling terms close to the superconducting transition temperature. Guided by the largely different temperature scales of CDW and superconducting order, we treat the CDW as an effective field: not as a free parameter of the GL free energy to be minimized for, but rather as an input informed by experiments. 
Importantly, our paper clarifies the role of additional point-group or time-reversal symmetry breaking of the charge density wave for different superconducting scenarios.

The rest of this manuscript is organized as follows. In Section~\ref{sec:classification}, we discuss the relevant symmetry to classify superconducting orders on the kagome lattice in the presence of a ($2\times2$) CDW, namely the extended point group $C_{6v}'''$. The classification builds upon earlier work~\cite{holbaek:2023} by including superconducting orders that are periodic in a $2 \times 2$ enlarged unit cell---relevant for the \textit{A}VS kagome materials. The possible charge-density-wave orders, extensively discussed elsewhere~\cite{wagner:2023,christensen:2022,christensen:2021,park:2021}, are summarized for completeness. We then distinguish two main scenarios discussed in Sections~\ref{sec:homogeneous} and~\ref{sec:modulated}: First, the dominant superconductivity is homogeneous with center of mass momentum $\ve{q}=\ve{0}$, in which case we distinguish the case of a single-component order parameter and a two-component order parameter, such as exemplified by a chiral $d$-wave order. Second, we discuss the possibility of a dominant PDW with $\ve{q} = \ve{M}_{i}$, which partially reproduces recently published results~\cite{yao:2024tmp}. At the end of each section, based on the derived lowest-order coupling terms, we discuss the impact of different charge density waves---whether isotropic or breaking rotational or time-reversal symmetry---on the superconducting state. We finish in Section \ref{sec:conclusions} with our conclusions.

\section{Symmetry classification and Ginzburg-Landau theory}\label{sec:classification}
\subsection{CDW-enlarged unit cell}
\begin{figure}
\centering
\includegraphics[width=0.95\linewidth]{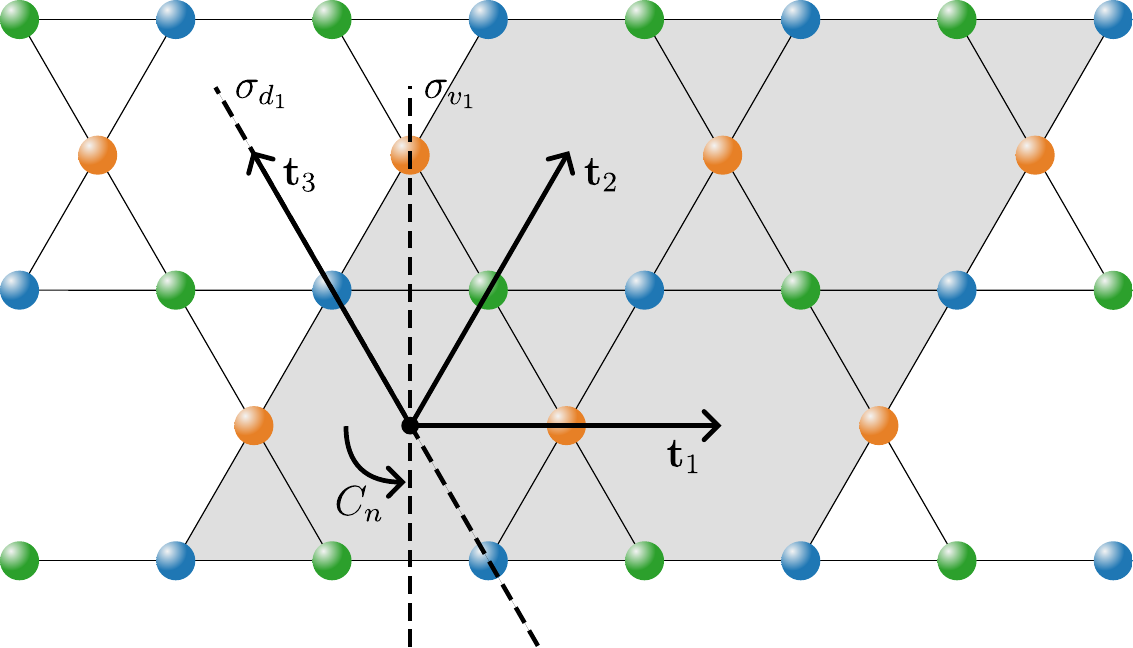}
\caption{The kagome lattice and the $2\times 2$ extended unit cell in grey. Also depicted are the symmetry operations of the extended point group $C_{6v}'''$ containing the primitive lattice vectors $\ve{t}_{i}$, along with planes for $\sigma_{v}$ and $\sigma_{d}$ mirror operations, and the (out-of-plane) axis of $C_{n}$ rotations.}
\label{fig:kagome_definitions}
\end{figure}

To derive a GL free energy, we start by discussing the relevant symmetry of the system. Considering, for simplicity, only a two-dimensional kagome layer and further neglecting spin-orbit coupling, we use $C_{6v}$ as the point group of the kagome plane. Since we are interested in commensurate order with a $2\times 2$ enlarged unit cell, we follow the scheme of Refs.~\onlinecite{venderbos:2016} and~\onlinecite{wagner:2023} and classify order parameters according to the extended point group $C_{6v}'''$. The extended point group contains the point group operations that leave the kagome lattice invariant and additional translations $t_{i}$, see Fig.~\ref{fig:kagome_definitions}, which do not map the enlarged unit cell onto itself~\cite{venderbos:2016}, thus allowing for orders that transform non-trivially under these translations.
In addition to the irreducible representations (irreps) $A_{1}$, $A_{2}$, $B_{1}$, $B_{2}$ (all one-dimensional) and $E_{1}$, $E_{2}$ (two-dimensional) inherited from $C_{6v}$, the $C_{6v}'''$ point group contains four three-dimensional irreducible representations $F_{n}$, $n=1,\ldots ,4$, which explicitly break the translational symmetry of the pristine kagome lattice~\footnote{Note that the assignment of mirrors $\sigma_{v}$ and $\sigma_{d}$ relative to the lattice is not unique. The convention used for these mirrors then fixes the irreps $B_{1}$, $B_{2}$, $F_{3}$ and $F_{4}$. Furthermore, in the notation of Ref.~\onlinecite{christensen:2022}, the $F_{n}$ irreps correspond to, respectively: $F_{1}\sim M_{1}^{+}$, $F_{2}\sim M_{2}^{+}$, $F_{3}\sim M_{4}^{-}$, and $F_{4}\sim M_{3}^{-}$.}. The character table of the extended point group $C_{6v}'''$ is given in App.~\ref{app:charactertable}.

We classify electronic orders corresponding to finite mean-field expectation values distinguishing particle-hole ($\braket{c^{\dagger}_{i\sigma}c^{\phantom{\dag}}_{j\sigma'}}$) and particle-particle ($\braket{c_{i\sigma}c_{j\sigma'}}$) channels, where $c^{\dag}_{i\sigma}$ creates an electron on site $i$ with spin $\sigma$. In particular, we ask what order-parameter symmetries can be realized given a range $r=|i-j|$. For the convenience of the later discussion, we first summarize the results of Ref.~\onlinecite{wagner:2023} on charge ordering.
For on-site (OS) charge order (CO), $i=j$, the resulting mean-field is purely real and can be interpreted as a local potential. The twelve sites within the enlarged unit cell can host twelve orders, with the permutation representation $\mathcal{P}^{\rm CO}_{\rm OS}$ of these orders decomposing into $\mathcal{P}^{\rm CO}_{\rm OS} = A_{1} \oplus E_{2} \oplus F_{1} \oplus F_{3} \oplus F_{4}$. Concretely, we find three translationally invariant intra-unit-cell orders and nine CDW-type orders~\cite{venderbos:2016, wagner:2023}.

For longer-ranged mean-field order, we distinguish between purely real or imaginary expectation values. The real orders lead to renormalized hopping amplitudes over equidistant ranges, and we refer to them as bond orders. The respective decomposition for nearest-neighbor (NN) orders reads $\mathcal{P}^{\rm CO}_{\rm NN} = A_{1} \oplus B_{1} \oplus E_{1} \oplus E_{2} \oplus 2F_{1} \oplus F_{2} \oplus 2F_{3} \oplus F_{4}$. The imaginary orders correspond to currents on the bonds, which we classify in terms of flux orders (FOs) on the plaquettes of the lattice to avoid overcounting orders that only differ by local gauge transformations~\cite{venderbos:2016}. The flux orders from a nearest-neighbor decomposition can realize the irreps $\mathcal{P}^{\rm FO}_{\rm NN} = 2A_{2}' \oplus B_{2}' \oplus 2F_{2}' \oplus F_{4}'$, where we use primes to denote the additional time-reversal-symmetry breaking in this type of order. Note, again, that bond orders transforming as an $F_{n}$ irrep describe real CDWs, which we refer to as rCDWs, while $F_{2}'$ or $F_{4}'$ are denoted as iCDWs (imaginary CDWs).
We primarily focus on the $F_{1}$ rCDW and $F_{2}'$ iCDW orders, which are consistent with experiments and have been the focus of most theoretical studies~\cite{ratcliff:2021, guo:2024, christensen:2021, christensen:2022, wagner:2023}.

\begin{figure}
\centering
\includegraphics{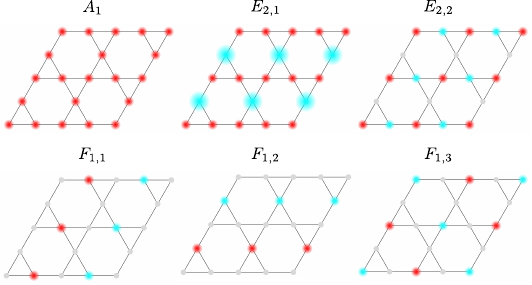}
\caption{Real-space representations of the on-site $A_{1}$, $E_{2}$ and $F_{1}$ spin-singlet pairing potentials. The color and its intensity at each site represent relative signs and strengths of the pairing potentials.}
\label{fig:OS_SC_orders}
\end{figure}

In the absence of spin-orbit coupling, the classification of superconducting orders reduces to a classification of the spatial degrees of freedom. In particular, the particle-particle mean-fields act as pairing potentials 
\begin{equation} 
    \mathcal{H}^{\rm mf} = \sum_{i,j}\Big[ \Delta^{\rm s/t}_{i,j} \big(c^{\dag}_{i\uparrow} c^{\dag}_{j\downarrow}\pm c^{\dag}_{j\uparrow} c^{\dag}_{i\downarrow}\big) + {\rm h.c.}\Big],
\end{equation}
and are either even under electron exchange and correspond to spin-singlet (s) pairing, or odd and correspond to spin-triplet (t) pairing. For an on-site order parameter, only spin-singlet orders are possible. The twelve sublattice sites of the $2 \times 2$ unit cell each come with a pairing potential $\Delta_{i}$. The permutation representation $\mathcal{P}^{\rm s}_{\rm OS}$ of these twelve potentials can be decomposed into the irreducible representations
\begin{equation}
    \mathcal{P}^{\rm s}_{\rm OS} = A_{1} \oplus E_{2} \oplus F_{1} \oplus F_{3} \oplus F_{4}.
\end{equation}
The pairing potentials corresponding to the trivial irrep $A_{1}$ are given by $\Delta_{i} = \Delta$ for all sites $i$. The $E_{2}$ irrep in the above decomposition does not break translational symmetry, but is rather modulated within the pristine kagome lattice unit cell, and has been found to appear within a recent fRG analysis~\cite{schwemmer:2024}. The irreps denoted by $F_{n}$, in contrast, correspond to translational-symmetry-breaking superconducting orders, in other words, they describe $\ve{q}=\ve{M}$ pair-density waves (PDWs). The real-space representations of the $A_{1}$, $E_{2}$ and $F_{1}$ pairing potentials are shown in Fig.~\ref{fig:OS_SC_orders}. A complete list of real-space representations is presented in App.~\ref{app:pairingpotentials}.

For nearest-neighbor pairing, both spin-singlet and spin-triplet order parameters are possible, with the former decomposing into
\begin{equation}
    \mathcal{P}^{\rm s}_{\rm NN} = A_{1} \oplus B_{1} \oplus E_{1} \oplus E_{2} \oplus 2F_{1} \oplus F_{2} \oplus 2F_{3} \oplus F_{4}.
\end{equation}
Here, $A_{1}$ is an extended $s$-wave order, and $E_{2}$ is the standard $d$-wave order. The $B_{1}$ and $E_{1}$ orders, which are odd under $\ve{k} \rightarrow -\ve{k}$ (in two dimensions, a $C_{2}$ rotation), at this range of interactions are enabled by the sublattice degree of freedom, and correspond to \textit{inter}band pairing states~\cite{holbaek:2023}. Finally, the spin-triplet order parameters decompose into
\begin{equation}
    \mathcal{P}^{\rm t}_{\rm NN} = A_{2} \oplus B_{2} \oplus E_{1} \oplus E_{2} \oplus F_{1} \oplus 2F_{2} \oplus F_{3} \oplus 2F_{4},
\end{equation}
where $B_2$ is an $f$-wave order, $E_1$ is $p$-wave, and $A_{2}$ and $E_{2}$ orders correspond again to \textit{inter}band pairing states at this range of interactions~\cite{holbaek:2023}. The pairing potentials of the $F_{n}$ irreps are PDW order parameters and are depicted in App.~\ref{app:pairingpotentials}. Note that for an $f$-wave pairing state of $B_{1}$ symmetry to appear, we have to consider next-to-nearest-neighbor interactions.

\subsection{Ginzburg-Landau free energy}
In this section, we discuss the general strategy for deriving the GL free energy. For this purpose, we first introduce the free energy for CDW order parameters as previously derived~\cite{christensen:2021, christensen:2022,wagner:2023}.
As mentioned above, the nearest-neighbor CDW order parameters can transform as any of the $F_{n}$ irreps for real bond orders, while a nearest-neighbor flux order, or iCDW, transforms as either $F_{2}'$ or $F_{4}'$.

\subsubsection{GL free energy for charge orders}
The main requirement for a  GL free energy is that it transforms as a scalar, in other words as $A_{1}$ in $C_{6v}'''$, and conserves time-reversal symmetry. A quadratic term in the free energy for a given irrep $\Gamma$ is thus allowed if the \textit{symmetrized} product of $\Gamma$, which we denote as $(\Gamma\otimes \Gamma)^{\rm S}$ in the following, contains $A_{1}$ in its decomposition~\cite{toledano:1987}. For the $F_{n}$ irreps, the symmetrized product decomposes into (see App.~\ref{app:charactertable} for the decomposition of the symmetrized higher powers of the $F_{n}$ irreps)
\begin{equation}\label{eq:cdw_square}
    (F_{n} \otimes F_{n})^{\rm S} = A_{1} \oplus E_{2} \oplus F_{1}.
\end{equation}
We denote the three components of a real CDW order parameter as $\ve{\rho} = (\rho_{1}, \rho_{2}, \rho_{3})^{T}$, where the subscript $i=1,2,3$ refers to momentum $\ve{M}_{i}$. The quadratic term, which, as is common, is assumed to be the only term with a temperature dependence~\cite{huang:2008}, has the form
\begin{equation}
    \mathcal{F}^{(2)}[\ve{\rho};T] = a_{\rm CDW}(T) \ve{\rho}^2,
\end{equation}
with $a_{\rm CDW}(T)$ becoming negative at the CDW transition temperature~\footnote{If the transition is first order, the transition happens before $a_{\rm CDW}(T)$ changes sign.}. For cubic terms, only the symmetrized third power of $F_{1}$ contains $A_{1}$, such that only this irrep allows for a term in the free energy proportional to $\rho_{1}\rho_{2}\rho_{3}$. The cubic term lifts the degeneracy between $3\ve{Q}$ CDW orders with $\text{sign}(\rho_{1}\rho_{2}\rho_{3}) < 0$ and $\text{sign}(\rho_{1}\rho_{2}\rho_{3}) > 0$, which for the isotropic case of $|\rho_{1}|=|\rho_{2}|=|\rho_{3}|$ are referred to as Star-of-David and tri-hexagonal bond orders, respectively~\cite{christensen:2021, wagner:2023}. The decomposition of the symmetrized fourth power is again the same for all $F_{n}$ irreps and contains $2A_{1}$. By projection of fourth-order monomials onto the $A_{1}$ irrep, the two invariant terms are obtained as $\ve{\rho}^{4}$ and $\rho_{1}^{2}\rho_{2}^{2} + \rho_{1}^{2}\rho_{3}^{2} + \rho_{2}^{2}\rho_{3}^{2}$.

Finally, note that when combining different order parameters with terms to $n$th and $m$th power, respectively, $A_{1}$ has to appear in the decomposition of the product of the \textit{symmetrized} $n$th and $m$th power of the respective irreps. As an example, with an $F_{2}'$ flux order (which we denote by $\ve{\rho}' = i(\rho_{1}',\rho_{2}',\rho_{3}')^{T}$ with $\rho_{i}'$ real) having the quadratic decomposition as given by Eq.~\eqref{eq:cdw_square}, we can construct a term from $(F_{2}' \otimes F_{2}')^{\rm S}$ transforming as $F_{1}$. This term can be combined with an $F_{1}$ CDW ($\ve{\rho}$) to yield an invariant cubic term in the GL free energy proportional to $\rho_{1}^{\phantom{'}} \rho_{2}'\rho_{3}' +$ cyclic permutations. This term, linear in the $F_{1}$ irrep, is an allowed coupling for any of the real or imaginary CDWs and promotes an induced $F_{1}$ CDW order, even when the leading CDW instability belongs to a different irrep. Concretely, even with $a_{\rm CDW} >0$ for $F_{1}$, a finite value $\rho_{1} \propto \rho_{2}' \rho_{3}' / a_{\rm CDW}$ appears (plus cyclic).

For concreteness, we discuss in the following four types of $F_{1}$ rCDW order parameters, each with different symmetry properties. There are two isotropic phases with $|\rho_{1}| = |\rho_{2}| = |\rho_{3}|$. These two orders, the Star-of-David [$\text{sign}(\rho_{1}\rho_{2}\rho_{3})<0$] and tri-hexagonal [$\text{sign}(\rho_{1}\rho_{2}\rho_{3})>0$] orders, are as mentioned distinguished by the third-order term in the GL free energy and preserve all $C_{6v}$ point group symmetries. Furthermore, we consider the phase represented by $\ve{\rho} = \pm(\rho,\rho+\delta, \rho)^{T}$ ($\delta \neq 0$) with point group $C_{2v}$, which spontaneously breaks $C_{3}$ rotational symmetry and we refer to as nematic. Finally, the phase sometimes referred to as (structurally) chiral has $|\rho_{1}|\neq |\rho_{2}|\neq |\rho_{3}|$ and breaks all mirror symmetries of $C_{6v}$ in addition to $C_{3}$ rotational symmetry~\footnote{Note that none of the orders considered break inversion, as they conserve $C_{2}$ and, in three dimensions, have an additional $z\mapsto -z$ mirror.}.

These rCDWs can be combined with iCDWs of $F_{2}'$ symmetry, additionally breaking time-reversal symmetry. Importantly, many of the combined orders carry a magnetic moment in the $z$ direction, $M_{z}$. Such a moment arises from combinations of an $F_{1}$ rCDW order with an $F_{2}'$ flux CDW~\cite{wagner:2023, tazai:2025}. Specifically, the moment can be written as~\footnote{The term $\propto \ve{\rho}\cdot \ve{\rho}'$ is also allowed for $F_{3}$ coupled with $F_{4}'$}
\begin{equation}\label{eq:magnetic_moment}
    M_{z} = m_{1} \ve{\rho}\cdot \ve{\rho}' + m_{2} \rho_{1}'\rho_{2}'\rho_{3}' + m_{3} (\rho_{1}\rho_{2}\rho_{3}' + {\rm cyclic}).
\end{equation}
Note that the phase discussed in Ref.~\onlinecite{xing:2024} corresponds here to the nematic rCDW solution with a flux phase of the form $\ve{\rho}' = i(\rho',0, -\rho')$ and has no magnetic moment.

\subsubsection{GL free energy for superconducting orders}
\begin{figure}
\centering
\includegraphics{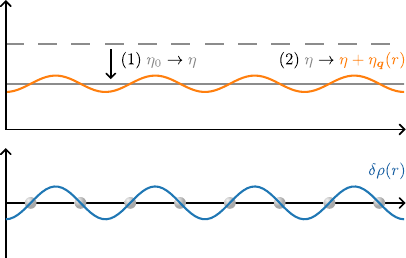}
\caption{Effect of a charge density wave $\delta \rho(r)$ on a homogeneous superconducting order parameter $\eta_{0}$: (1) direct coupling changes $T_{\rm c}$ and the absolute value of $\eta_{0}$ and (2) the order parameter is generically modulated with the same wave vector as the CDW.}
\label{fig:GL_illustration}
\end{figure}

When considering superconducting order parameters, the main difference to either the purely real or imaginary order parameters considered so far is that the superconducting order parameter is a complex quantity. Since the free energy has to be invariant under the $U(1)$ (gauge) symmetry, it cannot depend on the order parameter's phase. This puts the constraint that only an even number ($2n$) of superconducting order parameters can appear in the free energy. Further, an order parameter and its complex conjugate can be treated as separate order parameters and the decomposition of their product (as opposed to their \textit{symmetrized} product) has to contain $A_{1}$. Specifically, for an order parameter transforming according to the irrep $\Gamma$, a term in the free energy can be constructed if $A_{1}$ is in the decomposition of $(\otimes_{n}\Gamma)^{\rm S} \otimes (\otimes_{n}\Gamma^{*})^{\rm S}$. Based on these considerations, we construct in the next section the GL free energy for different homogeneous ($\ve{q}=\ve{0}$) superconducting order parameters, as well as their direct coupling to CDW orders and mutual coupling to PDW-type orders. For this purpose, we start with superconducting order parameters belonging to a one-dimensional irrep, such as (extended) $s$-wave order belonging to the $A_{1}$ irrep or spin-triplet $f$-wave order belonging to the $B_{2}$ irrep, before moving to the two-dimensional irrep $E_{2}$, which can realize a (chiral) $d$-wave order parameter. In both cases, we can ask (1) what the influence of the CDW on the (homogeneous) superconductivity is and (2) what kind of modulation in the form of a PDW order parameter can be induced, see illustration in Fig.~\ref{fig:GL_illustration}. In the subsequent section, we consider a PDW as the leading instability and consider its coupling to different $\ve{q}=\ve{0}$ order parameters through either the CDW or directly. As previously mentioned, we treat in both parts the CDW as an effective field interacting with the superconducting order, rather than a dynamical variable to be determined through minimization of the free energy.

\section{Dominant homogeneous superconducting order}\label{sec:homogeneous}
In this section, we assume the superconductivity emerges on the background of a CDW. Without the translation symmetry breaking of the CDW, the superconducting order parameter would be homogeneous, such that we can classify the order with respect to the irreps of $C_{6v}$. Our goal is to understand the effect of the CDW on the homogeneous superconducting order, as well as how the CDW induces spatially modulated superconductivity in the form of a PDW-type order parameter.

\subsection{One-dimensional irrep}
The free energy to fourth order in an order parameter, denoted by $\eta$, belonging to a one-dimensional irrep is given by
\begin{equation}
    \mathcal{F}[\eta; T] = a(T)|\eta|^{2} + b|\eta|^4,
\end{equation}
where $a(T) = a_{0} (T-T_{\rm c,0})$ changes sign at the superconducting transition temperature $T_{\rm c,0}$, and $a_{0}$ and $b$ are phenomenological parameters. Note that here and in the following, only the quadratic term of the dominant superconducting order parameter is assumed to have a temperature dependence, while all the other phenomenological parameters, including the quadratic term of the secondary orders, are temperature independent. For concreteness, we will here focus on order parameters belonging to $A_{1}$, but comment on differences to other irreps whenever applicable.

We can study how a CDW affects this order by looking at the GL free energy combining the two orders. While there are no terms to cubic order due to $U(1)$ and translational symmetry, we find a fourth-order---or bi-quadratic---term, allowed for all CDW orders, with the form
\begin{equation}
    \mathcal{F}^{(2;2)}[\eta; \ve{\rho}] = \beta |\eta|^2 \ve{\rho}^2.
\end{equation}
This term shifts the critical temperature of the superconductor, $T_{\rm c} = T_{\rm c,0} - \beta \ve{\rho}^2/a_{0}$. Most often, the presence of a CDW is associated with a decrease in $T_{\rm c}$ and conversely, $T_{\rm c}$ is increased when a CDW is suppressed, as seen also in \textit{A}VS~\cite{du:2021,wang:2021b,chen:2021,yuzki:2022,yuzki:2022b} or in some transition-metal dichalcogenides like 2$H$-TaS$_2$~\cite{kvashnin:2020b}. To derive the above term, we have used the fact that $(F_{n} \otimes F_{n})^{\rm S}$ contains $A_{1}$, see Eq.~\eqref{eq:cdw_square}. Given the decomposition of Eq.~\eqref{eq:cdw_square}, there is another allowed fourth-order term with homogeneous superconducting order parameters: coupling the order parameter to an order parameter of $E_{2}$ symmetry. Such a combination couples to a two-component vector transforming in the same way as uniaxial strain, namely as the $E_{2}$ irrep~\cite{christensen:2021,grandi:2023,xing:2024}, given by
\begin{equation}
    \ve{\rho}^{\rm ani.} = \begin{pmatrix}
        \rho_{2}^{2} - \frac12(\rho_{1}^{2} + \rho_{3}^{2})\\
        \frac{\sqrt{3}}{2}(\rho_{1}^{2} - \rho_{3}^{2})
    \end{pmatrix}.
\end{equation}
This term is non-zero for either the structurally chiral or nematic CDWs~\footnote{With the amplitude of one charge-density modulation different than the other two, three-fold rotational symmetry is broken and either $\ve{\rho}^{\rm ani.} \propto (1, 0)^{T}$ if $|\rho_{2}| \neq |\rho_{3}| = |\rho_{1}|$, $\ve{\rho}^{\rm ani.} \propto (1, -\sqrt{3})^{T}$ if $|\rho_{1}| \neq |\rho_{2}| = |\rho_{3}|$, or $\ve{\rho}^{\rm ani.} \propto (1, \sqrt{3})^{T}$ if $|\rho_{3}| \neq |\rho_{1}| = |\rho_{2}|$~\cite{xing:2024}. Other directions of $\ve{\rho}^{\rm ani.}$ are given by $|\rho_{1}| \neq |\rho_{2}| \neq |\rho_{3}|$.}. The resulting term in the free energy reads
\begin{equation}\label{eq:1D_homog_mixing}
    \mathcal{F}^{(1,1;2)}[\eta, \ve{\eta}_{E_2}; \ve{\rho}] = \frac{\mu}{2}\eta\ve{\eta}_{E_{2}}^{*}\cdot \ve{\rho}^{\rm ani.} + {\it c.c.},
\end{equation}
with the individual components of $\ve{\eta}_{E_{2}} = (\eta_{E_{2},1}, \eta_{E_{2},2})^{T}$ transforming as $(x^{2}-y^{2}, 2xy)^{T}$.
This term describes how an $A_{1}$ order parameter can induce an $E_{2}$ order parameter, which is time-reversal symmetric and parallel to $\ve{\rho}^{\rm ani.}$, and the term is only non-zero when the CDW components have different amplitudes.
In this sense, the term manifests the mixing of irreps due to the lowered symmetry of the system. Note that this coupling is allowed for all bond and flux orders, but only the one-dimensional irreps $A_{1}$ and $A_{2}$ couple to $E_{2}$ at this order \footnote{The one-dimensional irreps $B_{1}$ and $B_{2}$ mix with $E_{1}$ through the anisotropic order.}. In general, the symmetry-breaking of the CDW dictates what superconducting orders can be induced, as it sets the symmetry environment from which the superconducting phase emerges.

\begin{figure}
\centering
\includegraphics{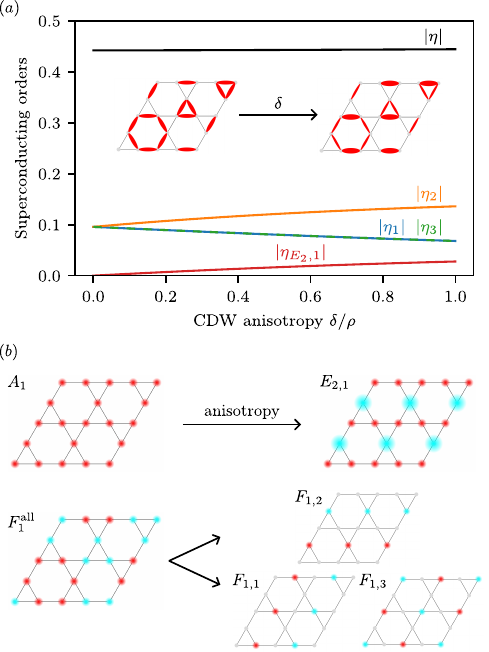}
\caption{(a) Illustration of the induced orders from a homogeneous superconducting order $|\eta|$ coexisting with a charge density wave $\bm{\rho} \propto (\rho, \rho + \delta, \rho)^{T}$ with $|\bm{\rho}| = 1$. The inset illustrates the tri-hexagonal CDW order that becomes anisotropic as $\delta$ increases. Figure obtained with $a(T) = -0.25$, $a_{E_{2},0} = a_{\rm PDW,0} = b = 1$, $\gamma = -0.75$, and $\mu = -0.25$. (b) Impact of the anisotropic CDW considered in (a) on the $A_{1}$ superconducting order: (1) inducing an $E_{2}$ component and (2) splitting the strengths of the induced PDW components.}
\label{fig:A1_CDW_induced_orders}
\end{figure}

Next, we study how the charge density wave induces a modulation of the superconducting order. We can capture this physics by including a pair-density-wave-type order parameter, which has three components $\ve{\eta} = (\eta_{1},  \eta_{2},  \eta_{3})^{T}$, see App.~\ref{app:pairingpotentials}, in our free energy. Here, subscript $i=1,2,3$ again refers to momentum $\ve{M}_{i}$. As we assume that this order is only induced, we use a quadratic term of the form
\begin{equation}\label{eq:PDW_subdominant2}
    \mathcal{F}^{(2)}[\ve{\eta}] = a_{\rm PDW} |\ve{\eta}|^{2},
\end{equation}
with $a_{\rm PDW}>0$ independent of temperature, signaling the leading instability in the one-dimensional order $\eta$. There is a third-order term coupling $\eta$ to the PDW through the (real) CDW,
\begin{equation}\label{eq:PDW_subdominant3}
    \mathcal{F}^{(1,1;1)}[\eta, \ve{\eta}; \ve{\rho}] = \frac{\gamma}{2} \sum_{i} \rho_{i} (\eta \eta_{i}^{*} + {\it c.c.}).
\end{equation}
\begin{table}[!tb]
\begin{tabular}{c|cccccc}
    & $A_{1}$ & $A_{2}$ & $B_{1}$ & $B_{2}$ & $E_{1}$ & $E_{2}$\\
    \hline
    $F_{1}$ & $F_{1}$ & $F_{2}$ & $F_{3}$ & $F_{4}$ & $F_{3(4)}$ & $F_{1(2)}$
\end{tabular}
\caption{Irrep of the induced pair-density-wave component given a $\ve{q}=\ve{0}$ superconducting order parameter and an $F_{1}$ CDW.}
\label{tab:induced_PDW}
\end{table}
For the case of an $A_{1}$ order parameter, the coupling induces a PDW that transforms according to the same irrep as the CDW, in other words, an $F_{n}$ CDW induces an $F_{n}$ PDW. The induced PDWs for the other $\ve{q}=\ve{0}$ irreps are summarized in Tab.~\ref{tab:induced_PDW} for an $F_{1}$ CDW. We emphasize that, although the order parameter for the modulated component resembles that of a pair density wave, the underlying mechanism differs. In the case of a primary homogeneous ($\bm{q} = \bm{0}$) superconducting order, the pair-density-wave component is induced by the spatial modulation due to the charge density wave---it represents a secondary order rather than the primary instability. Minimizing the free energy in Eqs.~\eqref{eq:PDW_subdominant2} and \eqref{eq:PDW_subdominant3} with respect to $\eta_{i}^{*}$, we find the induced PDW
\begin{equation}
    \eta_{i} = -\frac{\gamma \rho_{i}}{2 a_{\rm PDW}}\eta,
\end{equation}
such that, to this order, the induced PDW component is proportional to the CDW component with the same $\ve{q}$ vector. As a result, a nematic CDW will induce an anisotropic PDW that shares the same principal axis. In the case of a structurally chiral CDW with three inequivalent components, the induced PDW is also chiral with the same chirality. However, the chirality is \textit{not} generic, as we can see by examining the coupling to the next order in the CDW~\footnote{This term always couples to a PDW of $F_{1}$ symmetry. In other words, for a CDW that is not $F_{1}$, such as the flux CDW, both $F_{n}$ and $F_{1}$ PDWs can be induced. This is consistent with the fact that any $F_{n}$ CDW induces an $F_{1}$ CDW through a cubic term.}
\begin{equation}\label{eq:PDW_subdominant4}
    \mathcal{F}^{(1,1;2)}[\eta, \ve{\eta}; \ve{\rho}] = \frac{\nu}{2} [\rho_{2}\rho_{3} (\eta \eta_{1}^{*} + {\it c.c.}) + {\rm cyclic}].
\end{equation}
The induced PDW component for a primary CDW of $F_{1}$ symmetry reads 
\begin{equation}
    \eta_{1} = - \frac{\gamma \rho_{1} + \nu \rho_{2}\rho_{3}}{2 a_{\rm PDW}}\eta,
\end{equation}
and similarly for the other components. Consequently, the chirality can be reversed depending on the parameter values. Figure~\ref{fig:A1_CDW_induced_orders} summarizes the effect of a CDW on homogeneous superconductivity, showing the amplitude of induced orders as a function of CDW anisotropy $\delta$.

We further examine the coupling through a flux phase of either $F_{2}'$ or $F_{4}'$ symmetry. Since the CDW now changes sign under time-reversal symmetry, the coupling takes the form
\begin{equation}
    \mathcal{F}^{(1,1;1)}[\eta, \ve{\eta}; \ve{\rho}'] = i\frac{\gamma'}{2} \sum_{i} \rho_{i}' (\eta \eta_{i}^{*} - {\it c.c.}).
\end{equation}
Again, we find an induced PDW as above, but the induced component,
\begin{equation}
    \eta_{i} = - i \frac{\gamma' \rho_{i}'}{2 a_{\rm PDW}}\eta,
\end{equation}
has a phase shift of $\pm \pi/2$ with regards to the primary superconducting order parameter. This complex combination describes a time-reversal-symmetry-breaking superconducting state mirroring the time-reversal-symmetry breaking of the iCDW. Note that to lowest order, the induced order again transforms like the irrep of the CDW, such that here, we find an $F_{2}$ or $F_{4}$ PDW order parameter for a primary order of $A_{1}$ symmetry.

Finally, we comment on the situation, where both a rCDW and an iCDW are present. In this situation, the induced PDW components all have a relative phase to the $\ve{q}=0$ superconducting order, which is not fixed to $\pm \pi/2$ anymore. These non-trivial phases arise due to the combination of Eqs.~\eqref{eq:PDW_subdominant3} and \eqref{eq:PDW_subdominant4} with terms higher order in the CDW, such as
\begin{equation}
\begin{aligned}
    \mathcal{F}^{(1,1;1,1)}[\eta,\ve{\eta};\ve{\rho},\ve{\rho}'] & \propto i[(\rho_{2}\rho_{3}'-\rho_{2}'\rho_{3})(\eta\eta_{1}^{*} - c.c.)\\
    & + (\rho_{3}\rho_{1}'-\rho_{3}'\rho_{1})(\eta\eta_{2}^{*} - c.c.)\\
    & + (\rho_{1}\rho_{2}'-\rho_{1}'\rho_{2})(\eta\eta_{3}^{*} - c.c.)],
\end{aligned}
\end{equation}
which describes an additional coupling to an $F_{1}$ PDW order parameter.
Importantly, the non-trivial relative phase always leads to a superconducting order that breaks time-reversal symmetry, mimicking the time-reversal-symmetry breaking of the parent state with an iCDW.

\subsection{Two-dimensional irrep}
For an order parameter belonging to a two-dimensional irrep---we consider $E_{2}$ here---the free energy contains an additional fourth-order term,
\begin{multline}\label{eq:E2_free_energy}
    \mathcal{F}^{E_{2}}[\ve{\eta}_{E_{2}};T] = a_{E_{2}}(T) |\ve{\eta}_{E_{2}}|^{2} + b_{1} |\ve{\eta}_{E_{2}}|^{4}\\+ b_{2} (\eta_{E_{2},1}^{\phantom{*}} \eta_{E_{2},2}^{*} - \eta_{E_{2},1}^{*} \eta_{E_{2},2}^{\phantom{*}})^{2}.
\end{multline}
For $b_{2} > 0$, the ground state without a CDW is a chiral superconducting state, while $b_{2} < 0$ yields a nematic state breaking $C_{6}$ symmetry. Note that in this context, chiral does not refer to a structural chirality in the sense discussed above for the CDW, but rather a time-reversal-symmetry-breaking superconducting state such as $d+id$~\cite{kallin:2016}. We again in the following analyze the effect of a CDW on the homogeneous superconductivity before studying possible induced PDW states.

\begin{figure}
\centering
\includegraphics{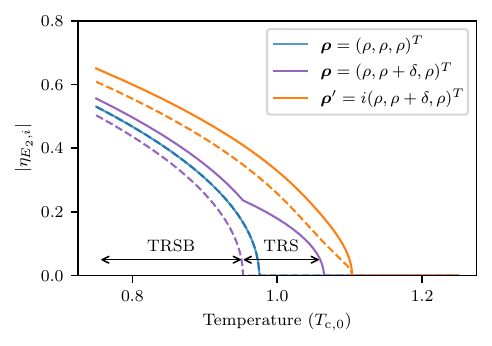}
\caption{The effect of CDW coupling on a two-component order parameter $E_{2}$ with $b_{2} > 0$ in Eq.~\eqref{eq:E2_free_energy}. Solid (dashed) lines indicate the component $|\eta_{E_{2},1}|$ ($|\eta_{E_{2},2}|$). For a rotationally symmetric real CDW (blue) with order parameter $\ve{\rho} = \rho(1,1,1)$, a $d+id$ order sets in at $T_{\rm c} < T_{\rm c,0}$. If the CDW breaks rotational symmetry (purple), $\ve{\rho} = (\rho, \rho+\delta, \rho)^{T}$, the transition is split and a time-reversal symmetric (TRS), anisotropic $d$-wave order is initially preferred. Only at a lower temperature, the second component appears with a phase shift of $\pi/2$. Finally, if the $3\ve{Q}$ CDW breaks three-fold rotational symmetry and is imaginary, the CDW again induces a TRSB solution at the transition. Parameters are chosen as $a_{E_{2},0} = T_{\rm c, 0} = b_{1} = 1$, $b_{2} = 0.8$, $\beta_{1} = 0.025$, $\beta_{2} = -0.5$, $\kappa m_{2} = -0.5$, $\delta/\rho = 2/7$, and $|\bm{\rho}| = 1$.}
\label{fig:E2_CDW_coupling}
\end{figure}

To second order in the CDW and $d$-wave order parameters, respectively, we find two possible direct coupling terms
\begin{equation}
\begin{aligned}
    \mathcal{F}^{(2;2)}[\ve{\eta}_{E_{2}}; \ve{\rho}]  & = \beta_{1} |\ve{\eta}_{E_{2}}|^{2} \ve{\rho}^{2}\\
    & +\beta_{2}\big[(|\eta_{E_{2},1}|^{2} - |\eta_{E_{2},2}|^{2})\rho^{\rm ani.}_{1} \\
    &-(\eta_{E_{2},1}^{\phantom{*}}\eta_{E_{2},2}^{*} + \eta_{E_{2},1}^{*}\eta_{E_{2},2}^{\phantom{*}})\rho^{\rm ani.}_{2}\big].
\end{aligned}
\end{equation}
This coupling is allowed for all of the CDWs, real or imaginary. As for a one-dimensional irrep, there is a trivial coupling of the CDW to the $E_{2}$ order parameter changing the superconducting $T_{\rm c}$. However, the second term is an additional coupling, only possible for a two-dimensional irrep such as $E_{2}$.
This term acts like a strain for an anisotropic charge density wave, lifting the degeneracy of the $E_{2}$ irrep. Consequently, the critical temperature is split and the system first enters a TRS, anisotropic state at the phase transition irrespective of the sign of $b_{2}$, see Fig.~\ref{fig:E2_CDW_coupling}. Finally, if the CDW components have different amplitude, a term of the form of Eq.~\eqref{eq:1D_homog_mixing} appears, which can induce another one-dimensional irrep.

Furthermore, unlike for one-dimensional irreducible representations, fourth-order couplings to two different CDW orders are allowed for the $E_{2}$ irrep. In particular, we find a coupling to an $F_{1}$ and $F_{2}$ CDW (also allowed for an $F_{3}$ and $F_{4}$ CDW), denoted respectively by $\ve{\rho}$ and $\tilde{\ve{\rho}}$, of the form
\begin{multline}
    \mathcal{F}^{(2;1,1)}[\ve{\eta}_{E_{2}}; \ve{\rho}, \tilde{\ve{\rho}}] = \beta_{3} [\frac{\sqrt{3}}{2}(|\eta_{E_{2},1}|^2 - |\eta_{E_{2}, 2}|^2) (\rho_{1}\tilde{\rho}_{1} - \rho_{3}\tilde{\rho}_{3})\\
    + (\eta_{E_{2},1}^{\phantom{*}}\eta_{E_{2},2}^{*} + \eta_{E_{2},1}^{*}\eta_{E_{2},2}^{\phantom{*}})(\rho_{2}\tilde{\rho}_{2} - \frac{1}{2}(\rho_{1}\tilde{\rho}_{1} + \rho_{3}\tilde{\rho}_{3}))].
\end{multline}
Note that any real combination of the $E_{2}$ components breaks ($C_{6v}$) point-group symmetries, explaining this type of coupling.

An interesting situation emerges for charge density waves that carry a magnetic moment $M_{z}$, see Eq.~\eqref{eq:magnetic_moment}, which couple to an $E_{2}$ order through
\begin{equation}
    \mathcal{F}^{(2;1,1)}[\ve{\eta}_{E_{2}}; \ve{\rho}, \ve{\rho}'] = i\kappa (\eta_{E_{2},1}^{\phantom{*}}\eta_{E_{2},2}^{*} - \eta_{E_{2},1}^{*}\eta_{E_{2},2}^{\phantom{*}}) M_{z}.
\end{equation}
There are two important consequences of this term: on the one hand, if both a real and flux CDW are present in such a way that $M_{z} \neq 0$, both components of the $E_{2}$ order will be present with a phase shift, in other words, the $E_{2}$ phase will break TRS irrespective of $b_{2}$, as one would expect in this situation, see Fig.~\ref{fig:E2_CDW_coupling}. On the other hand, if a real CDW is present and a chiral $d$-wave state condenses, a flux CDW will be induced.

Finally, we again study the modulation of the superconducting order by allowing for PDW-type order parameters. To lowest order, an $E_{2}$ superconducting order and an $F_{n}$ CDW order can induce an $F_{n}$ PDW component. Furthermore, an $E_{2}$ superconducting order can mix CDW and PDW irreps. $F_{1}$ and $F_{2}$ irreps can mix in the coupling to the $E_{2}$ order, and $F_{3}$ and $F_{4}$ irreps can mix. For the case of the two-dimensional irreps, the free energy thus contains terms coupling to two PDW irreps (see Tab.~\ref{tab:induced_PDW}),
\begin{multline}\label{eq:homo-to-PDW-F1}
    \mathcal{F}^{(1,1;1)}[\ve{\eta}_{E_{2}}, \ve{\eta}; \ve{\rho}] = \gamma_{1} \left\{ \left[\rho_{2}\eta_{2} - \frac{1}{2}(\rho_{1}\eta_{1} + \rho_{3}\eta_{3})\right]\eta_{E_{2},1}^{*}\right. \\+ \left.\frac{\sqrt{3}}{2}\left(\rho_{1}\eta_{1} - \rho_{3}\eta_{3}\right)\eta_{E_{2},2}^{*} + c.c.\right\}
\end{multline}
for a PDW of the same irrep as the CDW and
\begin{multline}\label{eq:homo-to-PDW-F2}
    \mathcal{F}^{(1,1;1)}[\ve{\eta}_{E_{2}}, \ve{\eta}; \ve{\rho}] = \gamma_{2} \left\{ \frac{\sqrt{3}}{2}\left(\rho_{3}\eta_{3} - \rho_{1}\eta_{1}\right)\eta_{E_{2},1}^{*}\right. \\+ \left.\left[\rho_{2}\eta_{2} - \frac{1}{2}(\rho_{1}\eta_{1} + \rho_{3}\eta_{3})\right]\eta_{E_{2},2}^{*} + c.c.\right\}
\end{multline}
for mixed-irrep CDW and PDW orders: given an $F_{1}$ CDW, this coupling leads to an $F_{2}$ PDW. As a result, the following PDW components are induced:
\begin{equation}
\begin{aligned}
    \eta_{1}^{F_{1}} & = \frac{\gamma_{1} \rho_{1}}{2a_{\rm PDW}^{F_{1}}}(\eta_{E_{2},1} - \sqrt{3} \eta_{E_{2},2}),\\
    \eta_{2}^{F_{1}} & = -\frac{\gamma_{1} \rho_{2}}{a_{\rm PDW}^{F_{1}}} \eta_{E_{2},1},\\
    \eta_{3}^{F_{1}} & = \frac{\gamma_{1} \rho_{3}}{2a_{\rm PDW}^{F_{1}}}(\eta_{E_{2},1} + \sqrt{3} \eta_{E_{2},2}).
\end{aligned}
\end{equation}
\begin{equation}
\begin{aligned}
    \eta_{1}^{F_2} & = \frac{\gamma_{2} \rho_{1}}{2a_{\rm PDW}^{F_{2}}}(\eta_{E_2,2} + \sqrt{3} \eta_{E_2,1}),\\
    \eta_{2}^{F_2} & = -\frac{\gamma_{2} \rho_{2}}{a_{\rm PDW}^{F_{2}}} \eta_{E_2,2},\\
    \eta_{3}^{F_2} & = \frac{\gamma_{2} \rho_{3}}{2a_{\rm PDW}^{F_{2}}}(\eta_{E_2,2} - \sqrt{3} \eta_{E_2,1}).
\end{aligned}
\end{equation}
A special situation is the case of the chiral $d$-wave state, $\ve{\eta}_{E_{2}} = |\ve{\eta}_{E_{2}}| (1, i)^{T}$, with an isotropic CDW $\ve{\rho}=\rho(1, 1, 1)^{T}$, which results in a $3\ve{Q}$ chiral PDW with $\eta_{j}^{F_{1}} = \gamma_{1}\rho |\ve{\eta}_{E_{2}}| / a_{\rm PDW} e^{-\frac{2\pi i}{3} (j - \frac{1}{2})}$ and $\eta_{j}^{F_{2}} \propto i\eta_{j}^{F_{1}}$.
Finally, for an $F_{2}'$ CDW $\ve{\rho}' = i\rho'(1,1,1)$, the couplings result in $F_{1}$ and $F_{2}$ PDWs, of the same form as above, but with a $\pi/2$ relative phase.

\begin{figure}
\centering
\includegraphics{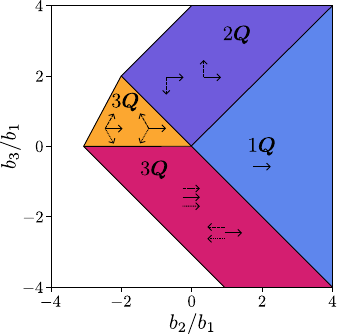}
    \caption{\label{fig:PDW_phase_diagram}Phase diagram of the pair density waves that minimize the free energy to fourth order with arrows denoting the relative phase structure of the three components. The free energy has been parameterized by $b_{2}/b_{1}$ and $b_{3}/b_{1}$. Note that the white region is where the free energy is unstable.}
\end{figure}

\section{Dominant pair-density wave}\label{sec:modulated}

We now turn to the situation, where a pair density wave with the same wave vector $\ve{q} = \ve{M}$ is dominant, in other words, the leading instability even without a CDW is the pair density wave. For this purpose, we first discuss the phase diagram of the PDW and how it will be influenced by the presence of a CDW. For completeness, we finish with a discussion of a pure PDW instability without a CDW and the induced order parameters in this situation.

\begin{figure}
\centering
\includegraphics{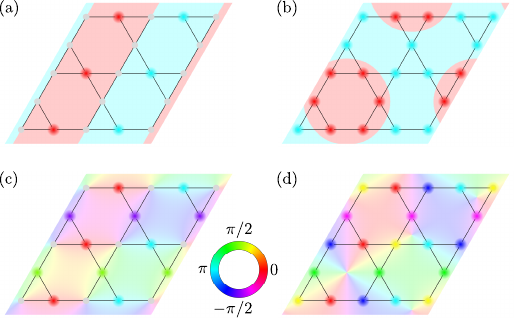}
\caption{Real-space illustration of the stable PDW phases of Fig.~\ref{fig:PDW_phase_diagram}, namely the (a) 1$\bm{Q}$, (b) 3$\bm{Q}$ TRS, (c) 2$\bm{Q}$ TRSB, and (d) 3$\bm{Q}$ chiral PDW order parameters. The color denotes the phase of the pairing potential.}
\label{fig:PDW_phases}
\end{figure}

\subsection{PDW in the presence of a CDW}
A PDW transforming as irrep $F_{n}$ ($n=1,2,3,4$) can be described using three complex order parameters $\ve{\eta} = (\eta_{1}, \eta_{2}, \eta_{3})^{T}$, see App.~\ref{app:pairingpotentials}.
The free energy for $\ve{\eta}$ has the general form
\begin{multline}\label{eq:free_PDW}
     \mathcal{F}^{\rm PDW}[\ve{\eta};T] = a_{\rm PDW}(T)|\ve{\eta}|^{2} + b_{1}|\ve{\eta}|^{4}\\
     + \frac{b_{2}}{2}\sum_{i\neq j}|\eta_{i}|^{2}|\eta_{j}|^{2} + \frac{b_{3}}{2}\sum_{i\neq j}\frac{1}{2}[\eta_{i}^{2}\eta_{j}^{*2} + c.c.],
\end{multline}
where $a_{\rm PDW}(T) = a_{\rm{PDW}, 0}(T - T_{\rm c,0})$ changes sign at the critical temperature $T_{\rm c,0}$ and $a_{\rm{PDW}, 0} > 0$ and the $b_{i}$ are temperature-independent phenomenological parameters. Figure~\ref{fig:PDW_phase_diagram} shows the phase diagram for the pure pair density wave~\cite{yao:2024tmp}. Similar to the case of an $E_{2}$ irrep, here $b_{3} > b_{2}$ leads to a time-reversal-symmetry-breaking state. Note that there are two TRSB pair-density-wave states, namely a chiral $3\ve{Q}$ state with $\ve{\eta} = |\ve{\eta}|(1, e^{i \omega}, e^{-i\omega})^{T}$ and $\omega = 2\pi/3$ or $\omega = \pi/3$ and a $2\ve{Q}$ state with $\ve{\eta} = |\ve{\eta}|(1, \pm i, 0)^{T}$. The latter solution becomes stable for large enough $b_{3}/b_{2}$. In addition, a $3\ve{Q}$ real PDW appears for $-b_{3} > b_{2}$ and a $1\ve{Q}$ state is realized when the $b_{2} > 0$ term dominates, with both orders time-reversal symmetric. The real space phase structure of the PDW order parameters is illustrated in Fig.~\ref{fig:PDW_phases}.

We next discuss how a CDW order setting in above the superconducting $T_{\rm c,0}$ can change the phase diagram. For the case of a CDW transforming as $F_1$, we find a coupling to third order,
\begin{equation}\label{eq:PDW_bond}
    \mathcal{F}^{(2;1)}[\ve{\eta}; \ve{\rho}] = \frac{\gamma_{\rm PDW}}{2}[\rho_{1}(\eta_{2}^{\phantom{*}}\eta_{3}^{*} + c.c.) + {\rm cyclic}].
\end{equation}
For an anisotropic CDW, we find an anisotropic PDW irrespective of the fourth-order terms in Eq.~\eqref{eq:free_PDW}, which becomes structurally chiral if the CDW is structurally chiral. However, as in the case of an induced PDW studied above, the chirality is not generic, as can be seen by introducing the fourth-order terms,
\begin{equation}\label{eq:PDW_bond_biquad}
\begin{aligned}
    \mathcal{F}^{(2;2)}[\ve{\eta}; \ve{\rho}] & = \beta_{1} \ve{\rho}^{2}|\ve{\eta}|^{2}\\
    & +\frac{\beta_{2}}{2} \sum_{i\neq j}\rho_{i}\rho_{j}\frac{1}{2}(\eta_{i}^{\phantom{*}}\eta_{j}^{*} + c.c.)\\
    & +\frac{\beta_{3}}{2} \sum_{i\neq j \neq k}[\rho_{i}^{2} - \frac{1}{2}(\rho_{j}^{2} + \rho_{k}^{2})]|\eta_{i}|^{2}.
\end{aligned}
\end{equation}
In particular, the second term can again invert the (structural) chirality of the PDW compared to the CDW.

Note that all the terms above involve the PDW to quadratic order, such that they in general determine both the critical temperature and the solution at $T_{\rm c}$. Only at lower temperatures, the fourth-order terms of Eq.~\eqref{eq:free_PDW} become dominant and might change the stable solution.

An interesting situation arises for an isotropic CDW, $\rho = |\rho_{1}|=|\rho_{2}|=|\rho_{3}|$: the $\beta_{3}$-term in Eq.~\eqref{eq:PDW_bond_biquad} vanishes, while the $\beta_{2}$-term together with the term in Eq.~\eqref{eq:PDW_bond} couple to the phase differences of the PDW components and can lead to frustration, in other words multiple terms that cannot be simultaneously optimized. To be specific, the terms $\propto (\eta_{i}^{\phantom{*}}\eta_{j}^{*} + c.c.)$ are individually minimized for relative phases of $0$ or $\pi$ in such a way that not all three phase relations can be fulfilled. The energy is instead minimized by having phase differences of $2\pi/3$. For a CDW that realizes a tri-hexagonal configuration, ${\rm sign}(\rho_{1}\rho_{2}\rho_{3})>0$, the phase dependent terms of Eqs.~\eqref{eq:PDW_bond} and \eqref{eq:PDW_bond_biquad} become frustrated for $\gamma_{\rm PDW} + \beta_{2} \rho > 0$, resulting in a chiral PDW with broken TRS. For the Star-of-David CDW, in contrast, ${\rm sign}(\rho_{1}\rho_{2}\rho_{3})<0$, and frustration happens for $\gamma_{\rm PDW} - \beta_{2} \rho < 0$.

\begin{figure}
\centering
\includegraphics{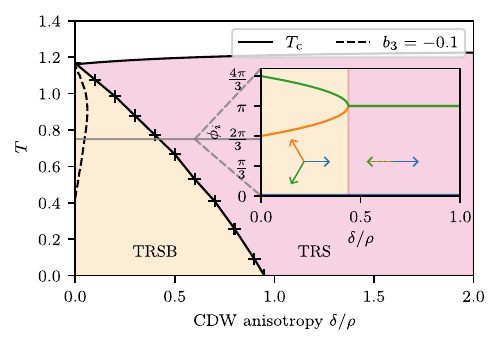}
\caption{\label{fig:Crossover_frustration}Phase diagram showing the transition from the frustrated regime with $\rho_{i} = \rho$ and $\gamma_{\rm PDW} + \beta_{2}\rho > 0$, where the $3\ve{Q}$ chiral PDW minimizes the free energy, to a nematic charge density wave, where a PDW with two phases equal to each other and with a $\pi$ difference to the last one is favored. Figure obtained with $a_{\rm PDW,0} = T_{\rm c,0} = b_{1} = -b_{2} = 1$, $b_{3} = \beta_{1} = 0$, $\gamma_{\rm PDW} = 0.5$, $\beta_{2} = 0.2$, $\beta_{3} = 0.1$, $\ve{\rho} \propto (\rho, \rho+\delta, \rho+\delta)^{T}$, and $|\bm{\rho}| = 1$. The dashed black line shows the TRSB-TRS transition for $b_{3}=-0.1$.}
\end{figure}

Figure~\ref{fig:Crossover_frustration} shows the phase diagram for the PDW when the isotropic CDW leads to frustration as a function of CDW anisotropy $\delta$. To not favor a chiral PDW over a real $3\ve{Q}$ PDW, we have used $b_3 = 0$, but set $b_2<0$ in order to promote a $3\ve{Q}$ phase, see Fig.~\ref{fig:PDW_phase_diagram}. While the critical temperature of the PDW generically changes when changing the CDW order parameter, we have further kept this change minimal by fixing the absolute value $|\bm{\rho}|$. The situation described in Fig.~\ref{fig:Crossover_frustration} resembles the case of a chiral two-component superconductor, where the anisotropy lifts the degeneracy at $T_{\rm c}$ and thus, the TRSB solution only appears at a temperature below the superconducting transition, see Fig.~\ref{fig:E2_CDW_coupling}. Note, again, that the chiral solution is not favored by the PDW free energy, but appears due to the frustration of the coupling to the CDW. Intriguingly, a chiral solution even persists for small anisotropy when $b_3 < 0$, in other words when the real $3\ve{Q}$ PDW dominates in the pristine phase diagram, Fig.~\ref{fig:PDW_phase_diagram}. In this special case, the chiral solution is only an intermediate phase that disappears again for small temperatures, where the quartic terms start to dominate.

The only other CDW order allowing for a third-order term in the free energy is the $F_{2}'$ flux order, where the additional term
\begin{equation}\label{eq:PDW_flux}
    \mathcal{F}^{(2;1)}[\ve{\eta}; \ve{\rho}'] =  \frac{\gamma_{\rm PDW}'}{2} [i\rho_{1}'(\eta_{2}^{\phantom{*}}\eta_{3}^{*} - c.c.) + {\rm cyclic}]
\end{equation}
is allowed. If only an iCDW is present, the term in Eq.~\eqref{eq:PDW_flux} has a similar effect as the phase-dependent terms of Eqs.~\eqref{eq:PDW_bond} and \eqref{eq:PDW_bond_biquad}, as illustrated in Fig.~\ref{fig:Crossover_frustration}: For an isotropic iCDW, Eq.~\eqref{eq:PDW_flux} leads to frustration and thus, to a TRS-breaking PDW, whereas sufficient anisotropy leads to a solution with phases of the form $0$, $\pi/2$, and $\pm \pi/2$.

Next, it is possible to couple the PDW to two different CDWs. We restrict our discussion here to the case of a $F_{1}$ rCDW and an $F_{2}'$ iCDW, where the free energy reads
\begin{equation}\label{eq:PDW_combinedCDW}
\begin{aligned}
    \mathcal{F}^{(2;1,1)}[\ve{\eta}; \ve{\rho}, \ve{\rho}'] & = i\frac{\beta_{4}}{2}[(\rho_{2}\rho_{3}' + \rho_{2}'\rho_{3})(\eta_{2}^{\phantom{*}}\eta_{3}^{*} - c.c.)\\
    & + (\rho_{3}\rho_{1}' + \rho_{3}'\rho_{1})(\eta_{3}^{\phantom{*}}\eta_{1}^{*} - c.c.)\\
    & + (\rho_{1}\rho_{2}' + \rho_{1}'\rho_{2})(\eta_{1}^{\phantom{*}}\eta_{2}^{*} - c.c.)].
\end{aligned}
\end{equation}
This term can be interpreted in two ways: first, it describes how a flux phase is induced in the case of a $3\ve{Q}$ chiral PDW and a $3\ve{Q}$ rCDW; and second, in the presence of both an $F_{1}$ rCDW and an $F_{2}'$ iCDW it affects the PDW solution. Taking into account all the terms that couple to the PDW phases, and considering the scenario where both an $F_{1}$ rCDW and an $F_{2}'$ iCDW are present---in which case the CDW orders will in general be anisotropic~\cite{christensen:2021, wagner:2023}---the phase differences between the PDW components are, in general, non-trivial. Specifically, they will deviate from values such as $0$, $\pi$, or $\pi/2$.

Finally, we note that a PDW induces homogeneous superconductivity through the CDW in the same way that a homogeneous superconducting order induces a PDW through the CDW, which is described in Eqs.~\eqref{eq:PDW_subdominant3}, \eqref{eq:homo-to-PDW-F1}, and \eqref{eq:homo-to-PDW-F2}. Importantly, a PDW in the presence of a CDW with the same wave vector will in general induce a superconducting order belonging to a one-dimensional irrep through Eq.~\eqref{eq:PDW_subdominant3}, thus opening a gap at the Fermi energy even for momenta not connected by $\ve{q}$. 
In particular, we find for the one-dimensional irrep
\begin{equation}
    \eta = -\frac{\gamma}{2 a} \ve{\rho}\cdot \ve{\eta},
\end{equation}
with $a>0$ the (temperature-independent) coefficient for the quadratic term of $\eta$. For completeness, $F_{1}$ or $F_{2}$ PDWs can induce two-dimensional $E_{2}$ homogeneous superconductivity, where the components take the form
\begin{equation}\label{eq:inducedE2}
\begin{aligned}
    &\eta_{E_{2},1} = -\frac{\gamma_{1}}{a_{E_{2}}}\left[\rho_{2}\eta_{2} - \frac{1}{2}(\rho_{1}\eta_{1} + \rho_{3}\eta_{3})\right]\\
    &\eta_{E_{2},2} = -\frac{\gamma_{1}}{a_{E_{2}}}\frac{\sqrt{3}}{2}(\rho_{1}\eta_{1} - \rho_{3}\eta_{3}),
\end{aligned}
\end{equation}
for a CDW and a PDW of the same irrep. The above expressions show that if at least one of either the PDW or the CDW is anisotropic, a homogeneous superconducting order belonging to $E_{2}$ is induced.
Note, however, that for the special case of an isotropic CDW with a $3\ve{Q}$ chiral PDW, no $A_{1}$ order is induced, since $\ve{\rho}\cdot\ve{\eta} = \rho|\ve{\eta}|(1 + e^{i2\pi/3} + e^{i4\pi/3}) = 0$ (the same conclusion holds if we include higher-order terms in the CDW), but Eq.~\eqref{eq:inducedE2} leads to an induced chiral $d$-wave order parameter.

\subsection{PDW as the only instability}
For completeness, we finish our considerations with the situation, where the leading instability is only towards a PDW. This situation has been discussed in Ref.~\onlinecite{agterberg:2011} and for \textit{A}VS in Ref.~\onlinecite{yao:2024tmp}.

First, we see from Eqs.~\eqref{eq:PDW_bond} and \eqref{eq:PDW_flux} directly that a PDW with at least two non-zero components induces a CDW, either a rCDW or an iCDW, depending on whether the PDW conserves time-reversal symmetry or not. As such, all the additional induced orders discussed above can be induced even when the PDW is the only leading instability. We can alternatively see the appearance of additional order parameters directly by considering the respective terms in the free energy~\cite{agterberg:2011}. In particular, we can derive all additional homogeneous superconducting order parameters from the decomposition of $F_{n} \otimes (F_{n}\otimes F_{n})^{\rm S}$. For concreteness, we only consider the $F_{1}$ PDW, which allows for a term in the free energy of the form
\begin{equation}
    \mathcal{F}^{(1,3)}[\eta,\ve{\eta}] = \lambda_{1}[\eta^{*}(\eta_{1}^{*}\eta_{2}^{\phantom{*}}\eta_{3}^{\phantom{*}} + {\rm cyclic}) + c.c.],
\end{equation}
leading to an induced $A_{1}$ order~\cite{zhou:2022}
\begin{equation}
    \eta = - \frac{\lambda_{1}}{a} (\eta_{1}^{*}\eta_{2}^{\phantom{*}}\eta_{3}^{\phantom{*}} + \eta_{1}^{\phantom{*}}\eta_{2}^{*}\eta_{3}^{\phantom{*}} + \eta_{1}^{\phantom{*}}\eta_{2}^{\phantom{*}}\eta_{3}^{*}).
\end{equation}
As expected, this term induces an $A_{1}$ order for a $3\ve{Q}$ PDW with a $0$ or $\pi$ phase difference between its components, but vanishes for the $3\ve{Q}$ chiral PDW with $\eta_{j}^{F_{1}} = |\ve{\eta}| e^{i2\pi j /3}$ and the PDWs, where at least one component is zero. Next, we find a term coupling to an order parameter of $E_{2}$ symmetry,
\begin{equation}
\begin{aligned}
    \mathcal{F}^{(1,3)}[\eta_{E_{2}},\ve{\eta}] & = \lambda_{2}\{[\eta_{2}^{*}\eta_{3}^{\phantom{*}}\eta_{1}^{\phantom{*}} - \frac{1}{2}(\eta_{3}^{*}\eta_{1}^{\phantom{*}}\eta_{2}^{\phantom{*}} + \eta_{1}^{*}\eta_{2}^{\phantom{*}}\eta_{3}^{\phantom{*}})]\eta_{E_{2},1}^{*}\\
    & + \frac{\sqrt{3}}{2}\eta_{2}(\eta_{3}^{\phantom{*}}\eta_{1}^{*} - \eta_{3}^{*}\eta_{1}^{\phantom{*}})\eta_{E_{2},2}^{*} + c.c.\},
\end{aligned}
\end{equation}
which leads to the $E_{2}$ components
\begin{equation}
\begin{aligned}
    \eta_{E_{2},1} & = - \frac{\lambda_{2}}{a_{E_{2}}}[\eta_{2}^{*}\eta_{3}^{\phantom{*}}\eta_{1}^{\phantom{*}} - \frac{1}{2}(\eta_{3}^{*}\eta_{1}^{\phantom{*}}\eta_{2}^{\phantom{*}} + \eta_{1}^{*}\eta_{2}^{\phantom{*}}\eta_{3}^{\phantom{*}})],\\
    \eta_{E_{2},2} & = - \frac{\lambda_{2}}{a_{E_{2}}}\frac{\sqrt{3}}{2}\eta_{2}(\eta_{3}^{\phantom{*}}\eta_{1}^{*} - \eta_{3}^{*}\eta_{1}^{\phantom{*}}).
\end{aligned}
\end{equation}
For a $3\ve{Q}$ chiral PDW, with $\eta_{j}^{F_{1}} = |\ve{\eta}| e^{i2\pi j /3}$, no $A_{1}$ is induced, while we find a chiral $E_{2}$ order parameter with $\ve{\eta}_{E_{2}} \propto (1, \pm i)^{T}$. For a real $3\ve{Q}$ PDW with $|\eta_{j}^{F_{1}}| = |\ve{\eta}|$, on the other hand, only an $A_1$ order parameter is induced.

\section{Conclusions}\label{sec:conclusions}
When superconductivity emerges out of a normal state already hosting a charge-density wave, the superconducting state adapts to the reduced symmetries of the host state. In this work, we have provided a comprehensive Ginzburg-Landau analysis for an $\ve{M}$ point CDW on the kagome lattice to study in detail, how the broken symmetries of the parent state influence the superconducting order parameter. In particular, the superconductivity develops a spatial modulation, which can be captured by a PDW-like order parameter. Furthermore, when the CDW breaks additional point group symmetries in the form of inequivalent CDW components---as was suggested for the members of the \textit{A}VS family---different irreps can mix. In addition, the PDW components in this situation again mimic the broken symmetries of the CDW in the form of an anisotropic modulation. Finally, in the case of a two-component order parameter or a PDW, anisotropy in the CDW can lead to a double transition, which could be experimentally observed.

It has been suggested that the charge-density-wave order in the kagome metals breaks time-reversal symmetry in the form of flux order. For a single-component superconductor, such as for $s$- or $f$-wave pairing, such time-reversal symmetry breaking leads to a combination of primary order parameter and induced pair-density wave with a non-trivial relative phase. Only in the case of a multi-component superconductor or a dominant PDW, the superconducting state may adapt to the iCDW through a chiral order.

An intriguing situation can arise for a PDW on the background of an isotropic CDW, where time-reversal symmetry can be spontaneously broken in the form of a chiral $3\ve{Q}$ PDW due to frustration. Unlike the cases with $\ve{q}=0$ order, this case leads to an additional reduction in the symmetry compared to the normal state, with possible experimental signatures in $\mu$SR, to name an example.

While our analysis provides qualitative insights into the effect of a CDW on the superconducting state, it cannot quantitatively capture the influence a CDW has on the superconducting state. However, it can provide guidance to microscopic calculations that can give more quantitative insight. Given the usually detrimental effect of TRSB on superconductivity, a particularly interesting question in this respect is the stability of various superconducting pairing channels against flux order.

\begin{acknowledgments}
We acknowledge insightful discussions with Brian M.~Andersen, Morten H.~Christensen, Fernando de Juan, Andreas Kreisel, Hiroaki Kusunose, Bernhard L\"uscher, Titus Neupert, Manfred Sigrist, Glenn Wagner, and Jia-Xin Yin.
M.~H.~F and S.~C.~H are supported by the Swiss National Science Foundation (SNSF) through Division II (number 207908). S.~C.~H. further acknowledges support from the Swiss National Science Foundation (project 200021E\_198011) as part of the FOR 5249 (QUAST) led by the Deutsche Forschungsgemeinschaft (DFG, German Research Foundation).
\end{acknowledgments}

\begin{widetext}
\appendix

\section{Extended point group \texorpdfstring{$C_{6v}'''$}{C6v'''}}\label{app:charactertable}
In this appendix, we provide the character table for the extended point group $C_{6v}'''$, Tab.~\ref{tab:charactertable}, as well as the decomposition of the symmetrized products for the three-dimensional irreps, Tab.~\ref{tab:symmetrizedproduct}. The full product table for the irreps of $C_{6v}'''$ can be found, e.g., in Ref.~\onlinecite{wagner:2023}.
\begin{table}[h!]
\begin{center}
\begin{tabular}{ccccccccccc}
& $I$ & $t_{i}$ & $C_{2}$ & $t_{i}C_{2}$ & $C_{3}$ & $C_{6}$ & $\sigma_{v}$ & $t_{i}\sigma_{v}$ & $\sigma_{d}$ & $t_{i}\sigma_{d}$\\
\hline
$|\mathcal{C}|$ & $1$ & $3$ & $1$ & $3$ & $8$ & $8$ & $6$ & $6$ & $6$ & $6$\\
\hline
$A_{1}$ & $1$ & $1$ & $1$ & $1$ & $1$ & $1$ & $1$ & $1$ & $1$ & $1$\\
$A_{2}$ & $1$ & $1$ & $1$ & $1$ & $1$ & $1$ & $-1$ & $-1$ & $-1$ & $-1$\\
$B_{1}$ & $1$ & $1$ & $-1$ & $-1$ & $1$ & $-1$ & $1$ & $1$ & $-1$ & $-1$\\
$B_{2}$ & $1$ & $1$ & $-1$ & $-1$ & $1$ & $-1$ & $-1$ & $-1$ & $1$ & $1$\\
\hline
$E_{1}$ & $2$ & $2$ & $-2$ & $-2$ & $-1$ & $1$ & $0$ & $0$ & $0$ & $0$\\
$E_{2}$ & $2$ & $2$ & $2$ & $2$ & $-1$ & $-1$ & $0$ & $0$ & $0$ & $0$\\
\hline
$F_{1}$ & $3$ & $-1$ & $3$ & $-1$ & $0$ & $0$ & $1$ & $-1$ & $1$ & $-1$\\
$F_{2}$ & $3$ & $-1$ & $3$ & $-1$ & $0$ & $0$ & $-1$ & $1$ & $-1$ & $1$\\
$F_{3}$ & $3$ & $-1$ & $-3$ & $1$ & $0$ & $0$ & $1$ & $-1$ & $-1$ & $1$\\
$F_{4}$ & $3$ & $-1$ & $-3$ & $1$ & $0$ & $0$ & $-1$ & $1$ & $1$ & $-1$
\end{tabular}
\end{center}
\caption{\label{tab:charactertable}Character table of the extended point group $C_{6v}'''$~\cite{venderbos:2016,wagner:2023}. In addition to the irreps of $C_{6v}$, $C_{6v}'''$ contains four three-dimensional irreps $F_{n}$ $(n = 1,2,3,4)$ that transform non-trivially under translations $t_{i}$ $(i=1,2,3)$. The one-, two-, and three-dimensional irreps are separated by horizontal lines.}
\end{table}

\begin{table}[h!]
\begin{center}
\begin{tabular}{c|c|c|c}
& $(\otimes_{2}\Gamma)^{\rm S}$ & $(\otimes_{3}\Gamma)^{\rm S}$ & $(\otimes_{4}\Gamma)^{\rm S}$\\
\hline
$F_{1}$ & $A_{1} \oplus E_{2} \oplus F_{1}$ & $A_{1} \oplus 2F_{1} \oplus F_{2}$ & $2A_{1} \oplus 2E_{2} \oplus 2F_{1} \oplus F_{2}$\\
$F_{2}$ & $A_{1} \oplus E_{2} \oplus F_{1}$ & $A_{2} \oplus F_{1} \oplus 2F_{2}$ & $2A_{1} \oplus 2E_{2} \oplus 2F_{1} \oplus F_{2}$\\
$F_{3}$ & $A_{1} \oplus E_{2} \oplus F_{1}$ & $B_{1} \oplus 2F_{3} \oplus F_{4}$ & $2A_{1} \oplus 2E_{2} \oplus 2F_{1} \oplus F_{2}$\\
$F_{4}$ & $A_{1} \oplus E_{2} \oplus F_{1}$ & $B_{2} \oplus F_{3} \oplus 2F_{4}$ & $2A_{1} \oplus 2E_{2} \oplus 2F_{1} \oplus F_{2}$
\end{tabular}
\end{center}
\caption{\label{tab:symmetrizedproduct}Decomposition of the symmetrized second, third and fourth power of irreducible representation $\Gamma = F_{n}$ with $n=1,2,3,4$ of $C_{6v}'''$.}
\end{table}

\section{Visualization of pair-density-wave orders}\label{app:pairingpotentials}
In Tabs.~\ref{tab:OS_SC_OP} - \ref{tab:NN_triplet_SC_OP}, we show the real-space structure of PDW order parameters for spin-singlet pairing with on-site (Tab.~\ref{tab:OS_SC_OP}) and nearest-neighbor (Tab.~\ref{tab:NN_SC_OP}) structure, as well as spin-triplet pairing with nearest-neighbor structure (Tab.~\ref{tab:NN_triplet_SC_OP}).
To illustrate how to obtain a Hamiltonian from these illustrations, we consider the simplest case of an on-site $F_{1}$ superconducting order, see first row of Tab.~\ref{tab:OS_SC_OP}. The real-space Hamiltonian can be written as
\begin{equation}
    \mathcal{H}_{\rm SC}^{F_{1}} = \sum_{i} \eta_{1}\cos(\ve{M}_{1}\cdot\ve{R}_{i})c_{A\ve{R}_{i}\uparrow}^{\dag}c_{A\ve{R}_{i}\downarrow}^{\dag} + \eta_{2}\cos(\ve{M}_{2}\cdot\ve{R}_{i})c_{B\ve{R}_{i}\uparrow}^{\dagger}c_{B\ve{R}_{i}\downarrow}^{\dagger} - \eta_{3}\cos(\ve{M}_{3}\cdot\ve{R}_{i})c_{C\ve{R}_{i}\uparrow}^{\dagger}c_{C\ve{R}_{i}\downarrow}^{\dagger} + h.c.,
\end{equation}
where the sum is over the unit cells of the pristine kagome lattice, and the cosine accounts for the modulation of the superconducting mean field. The prefactors $\eta_{i}$ correspond to those in the Ginzburg-Landau theory. Here, we defined the $\ve{M}$ vectors
\begin{equation}
    \ve{M}_{1} = \frac{1}{2}\ve{g}_{1} \equiv \pi(1, -\frac{1}{\sqrt{3}})^{T},\quad \ve{M}_{2} = \frac{1}{2}\ve{g}_{2} \equiv \pi(0, \frac{2}{\sqrt{3}})^{T},\quad \ve{M}_{3} \equiv -\frac{1}{2}(\ve{g}_{1} + \ve{g}_{2}).
\end{equation}
After applying a Fourier transform and band folding, we arrive at the Hamiltonian
\begin{equation}
    \mathcal{H}_{\text{SC}}^{F_{1}} = \frac{1}{4N_{\ve{k}}}\sum_{\ve{k}\in\text{BZ}} (\Psi_{\ve{k}\uparrow}^{\dagger})^{T} \begin{pmatrix}
        0 & D_{1} & D_{2} & D_{3}\\
        D_{1} & 0 & D_{3} & D_{2}\\
        D_{2} & D_{3} & 0 & D_{1}\\
        D_{3} & D_{2} & D_{1} & 0
    \end{pmatrix}\Psi_{-\ve{k}\downarrow}^{\dagger} + h.c.,
\end{equation}
expressed in the basis $\Psi_{\ve{k}\sigma}^{\dagger} = (\psi_{\ve{k}\sigma}^{\dagger}, \psi_{\ve{k} + \ve{M}_{1}\sigma}^{\dagger}, \psi_{\ve{k} + \ve{M}_{2}\sigma}^{\dagger}, \psi_{\ve{k} + \ve{M}_{3}\sigma}^{\dagger})^{T}$, where $\psi_{\ve{k}\sigma}^{\dagger} = (c_{A\ve{k}\sigma}^{\dagger}, c_{B\ve{k}\sigma}^{\dagger}, c_{C\ve{k}\sigma}^{\dagger})^{T}$. The $3\times 3$ matrices $D_{i}$ are given by
\begin{equation}
    D_{1} = \eta_{1}\begin{pmatrix}
        1 & 0 & 0\\
        0 & 0 & 0\\
        0 & 0 & 0
    \end{pmatrix},\quad D_{2} = \eta_{2}\begin{pmatrix}
        0 & 0 & 0\\
        0 & 1 & 0\\
        0 & 0 & 0
    \end{pmatrix},\quad D_{3} = \eta_{3}\begin{pmatrix}
        0 & 0 & 0\\
        0 & 0 & 0\\
        0 & 0 & -1
    \end{pmatrix}.
\end{equation}

\begin{table}[h!]
\begin{center}
\resizebox{0.9\columnwidth}{!}{
\begin{tabular}{ccccc}
\toprule
\textbf{Irrep} & \multicolumn{3}{c}{\textbf{Components}} & \textbf{Symmetric superposition}\\
\hline
$F_{1}$ & $F_{1,1}$ & $F_{1,2}$ & $F_{1,3}$ &\\
& \includegraphics[width=3cm]{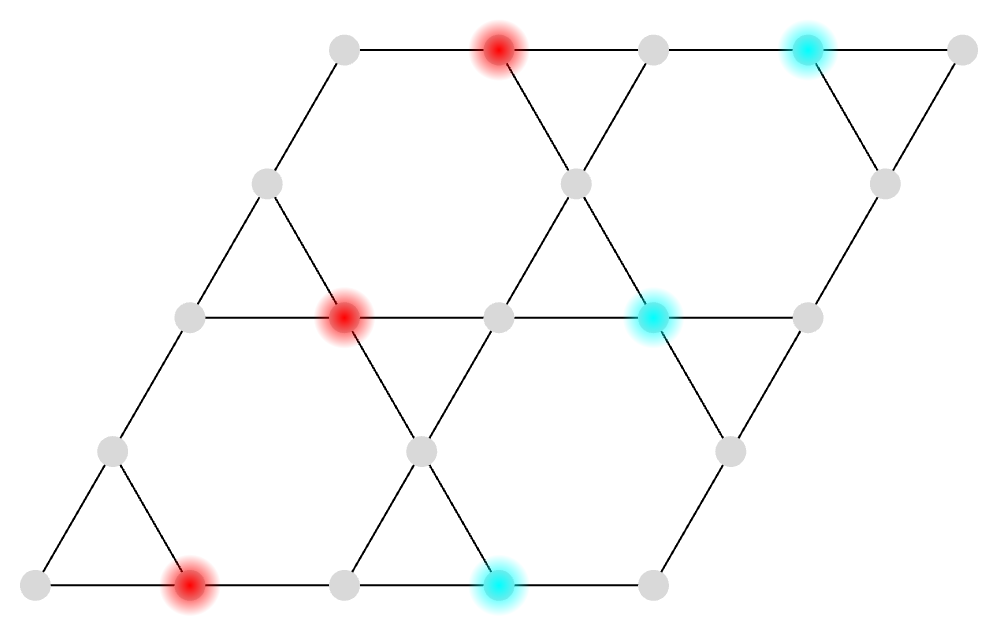} & \includegraphics[width=3cm]{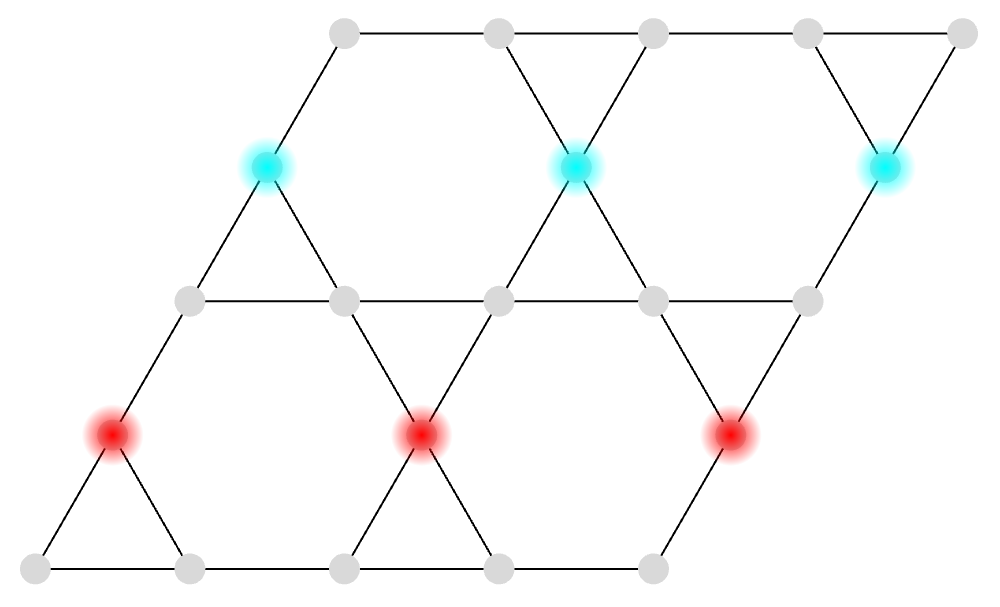} & \includegraphics[width=3cm]{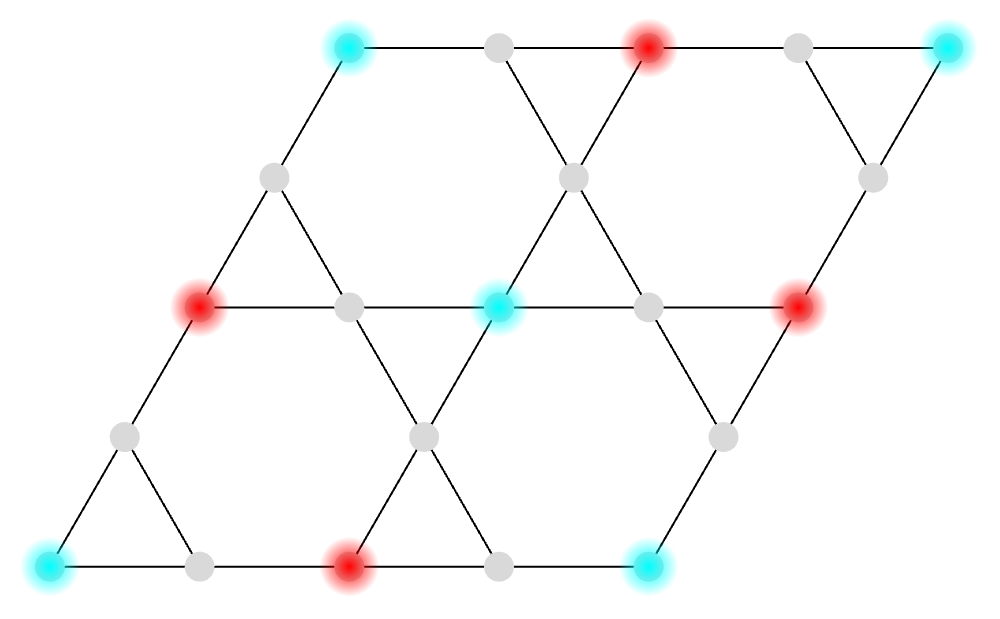} & \includegraphics[width=3cm]{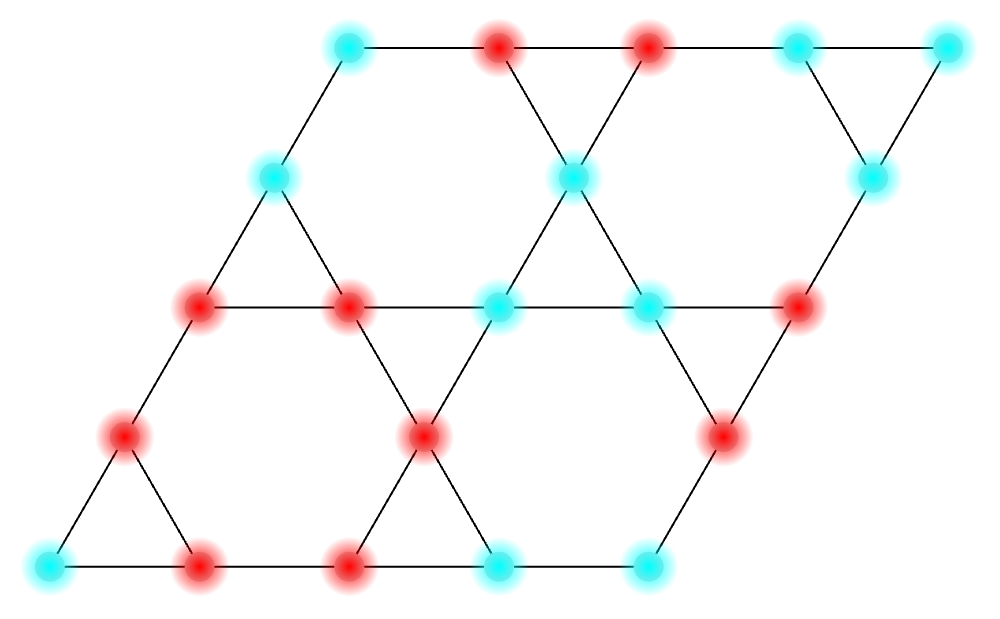}\\
$F_{3}$ & $F_{3,1}$ & $F_{3,2}$ & $F_{3,3}$ &\\
& \includegraphics[width=3cm]{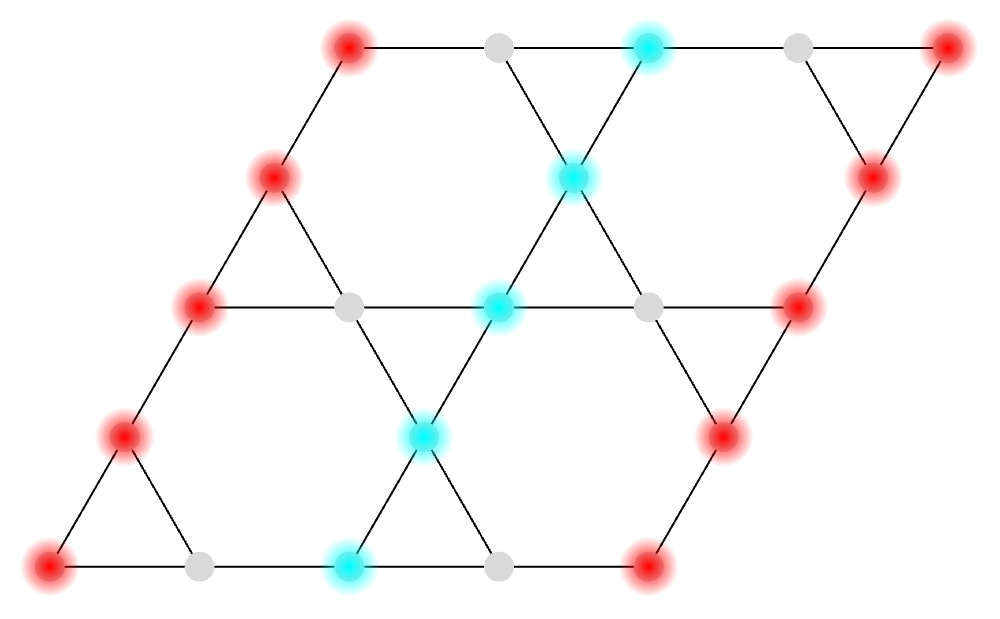} & \includegraphics[width=3cm]{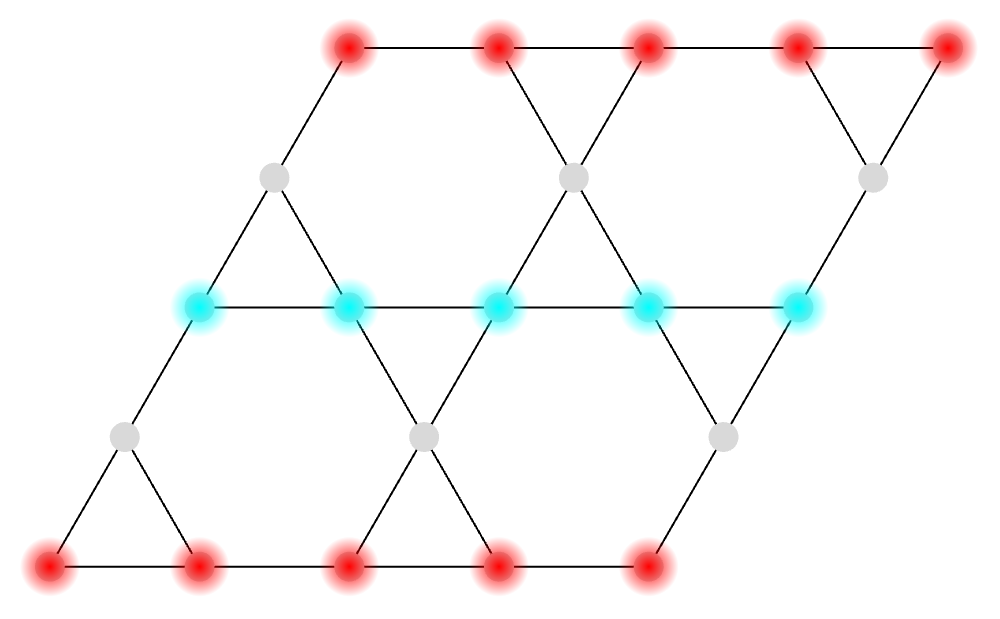} & \includegraphics[width=3cm]{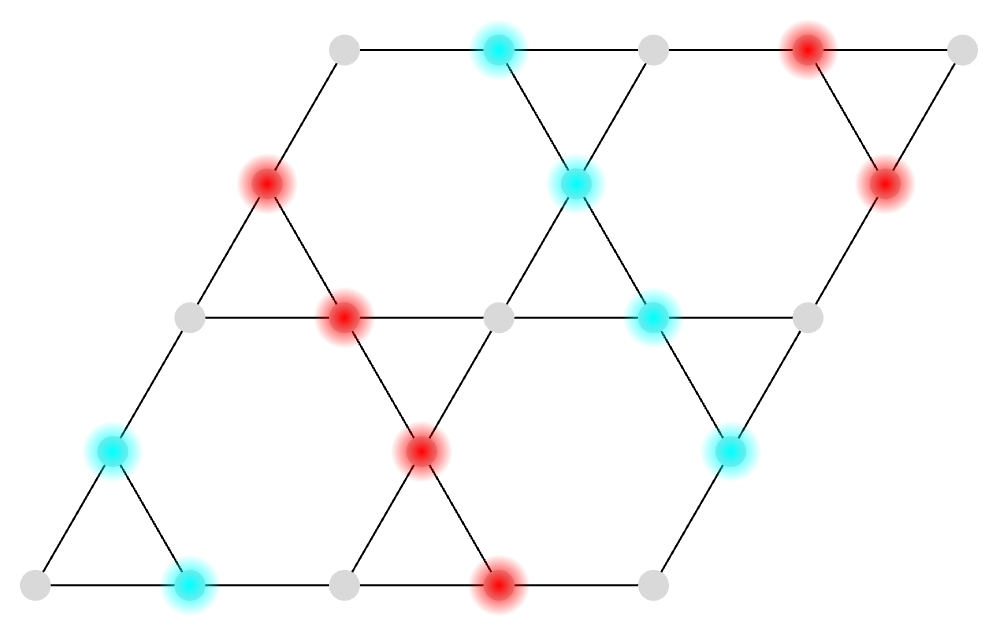} & \includegraphics[width=3cm]{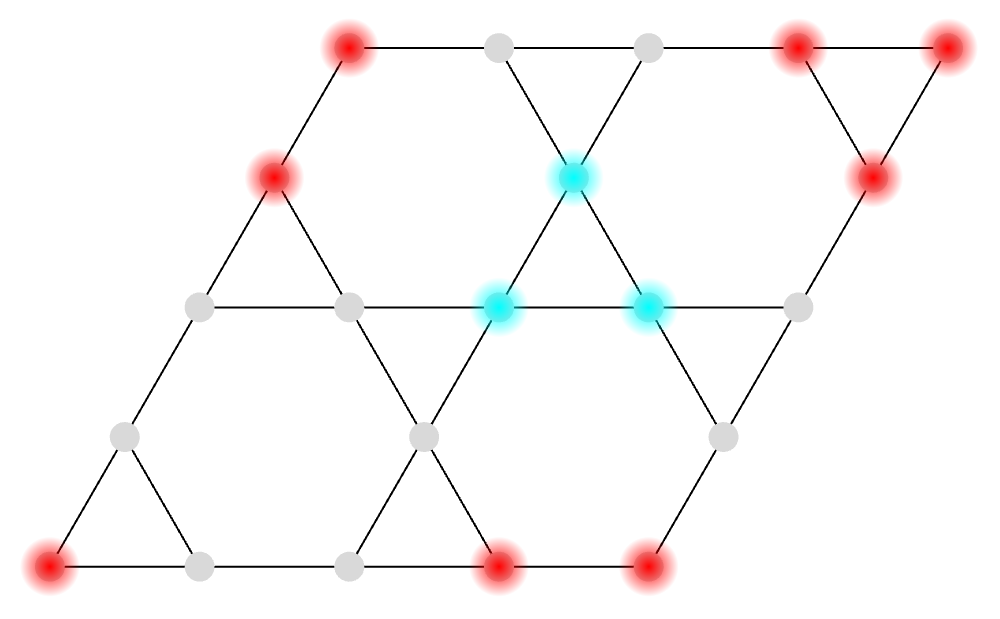}\\
$F_{4}$ & $F_{4,1}$ & $F_{4,2}$ & $F_{4,3}$ &\\
& \includegraphics[width=3cm]{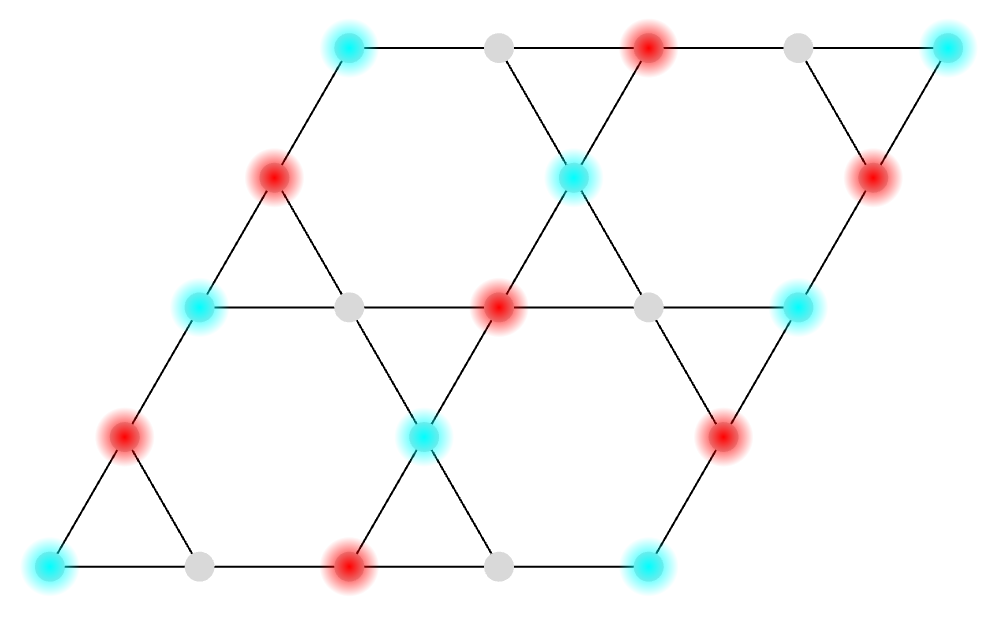} & \includegraphics[width=3cm]{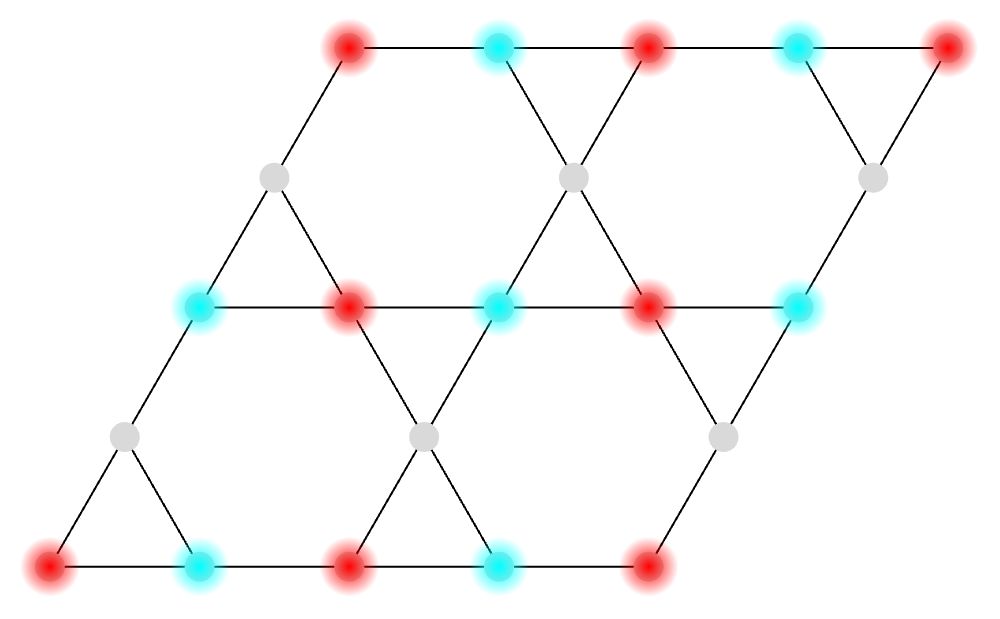} & \includegraphics[width=3cm]{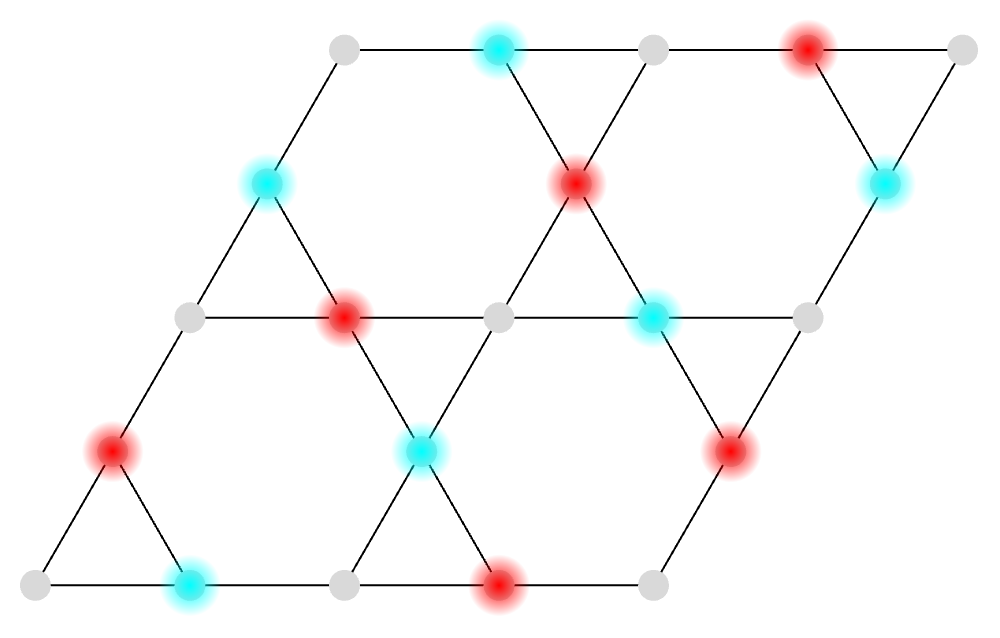} & \includegraphics[width=3cm]{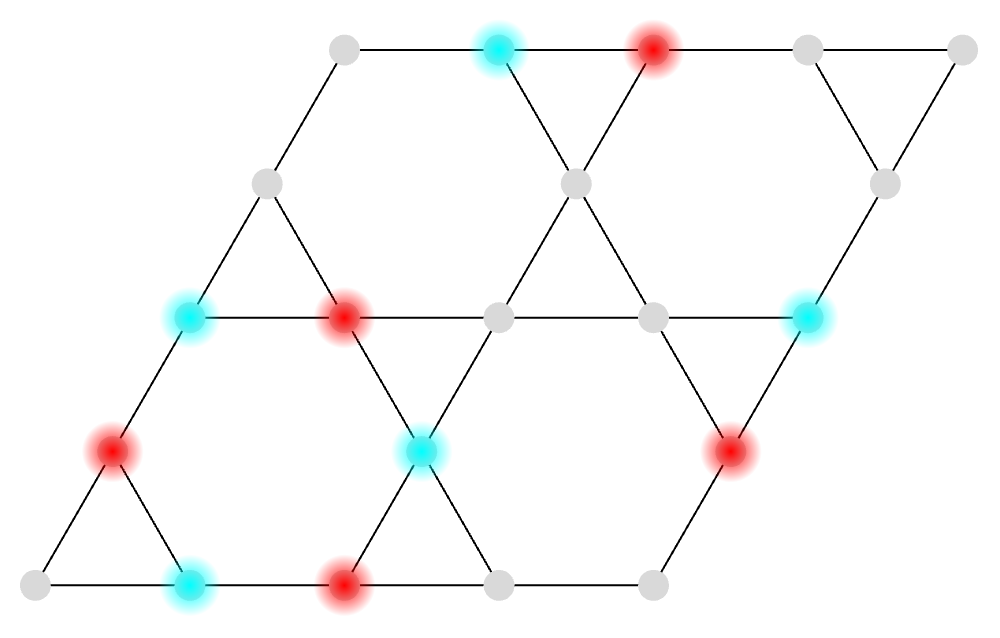}\\
\bottomrule
\end{tabular}}
\caption{\label{tab:OS_SC_OP}Real space illustrations of on-site pairing states on the kagome lattice which break translational symmetry. The color (red/blue) denotes the sign ($+1$/$-1$) of the on-site singlet pairing interaction. The first column provides the three-dimensional irrep, denoted according to the $C_{6v}'''$ point group. The second column shows each component of the irrep, and the last column shows a symmetric superposition of the three components.}
\end{center}
\end{table}

\begin{table}[h!]
\begin{center}
\resizebox{0.9\columnwidth}{!}{
\begin{tabular}{ccccc}
\toprule
\textbf{Irrep} & \multicolumn{3}{c}{\textbf{Components}} & \textbf{Symmetric superposition}\\
\hline
$F_{1}^{(1)}$ & $F_{1,1}^{(1)}$ & $F_{1,2}^{(1)}$ & $F_{1,3}^{(1)}$ &\\
& \includegraphics[width=3cm]{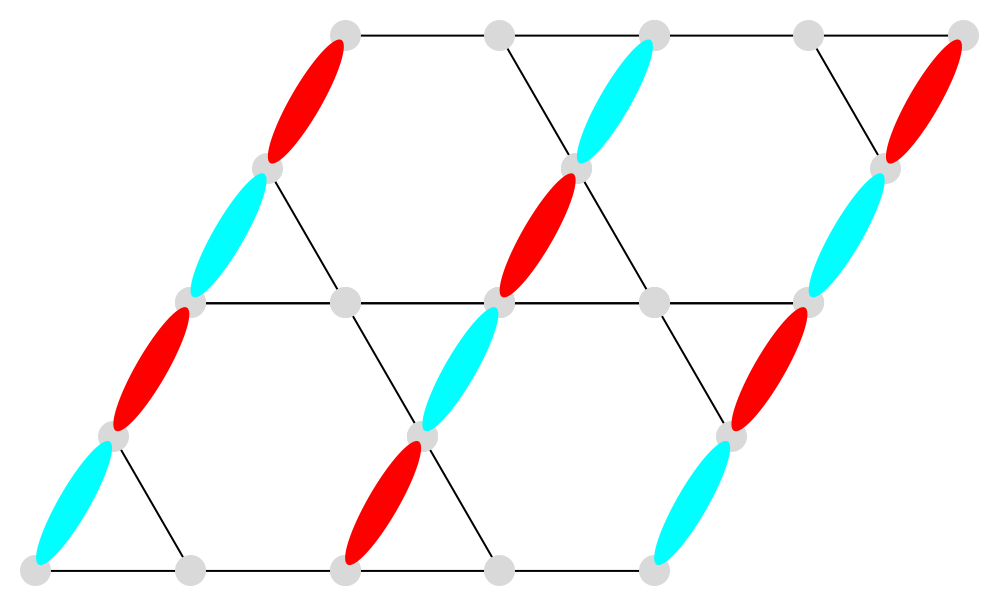} & \includegraphics[width=3cm]{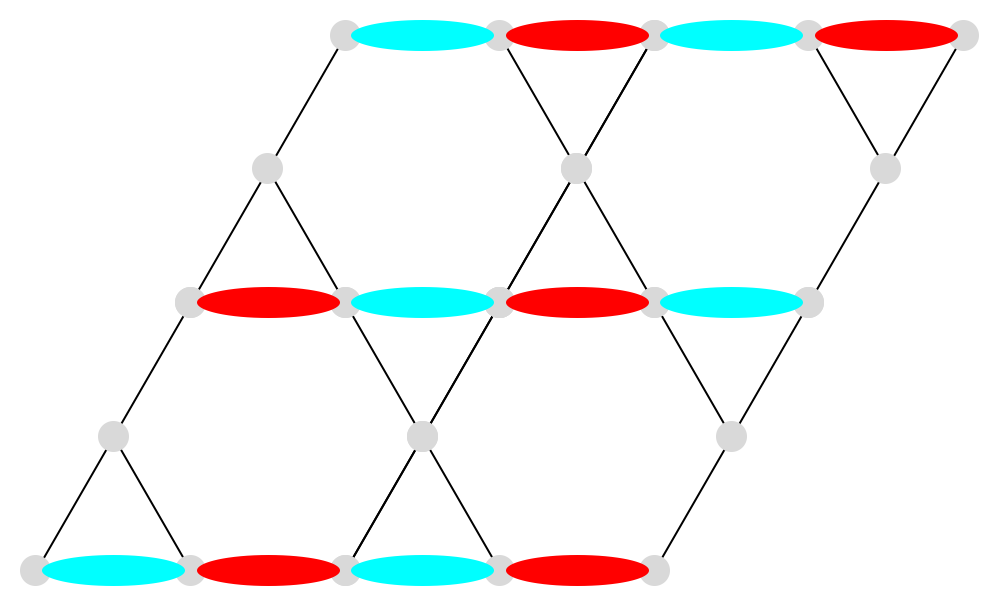} & \includegraphics[width=3cm]{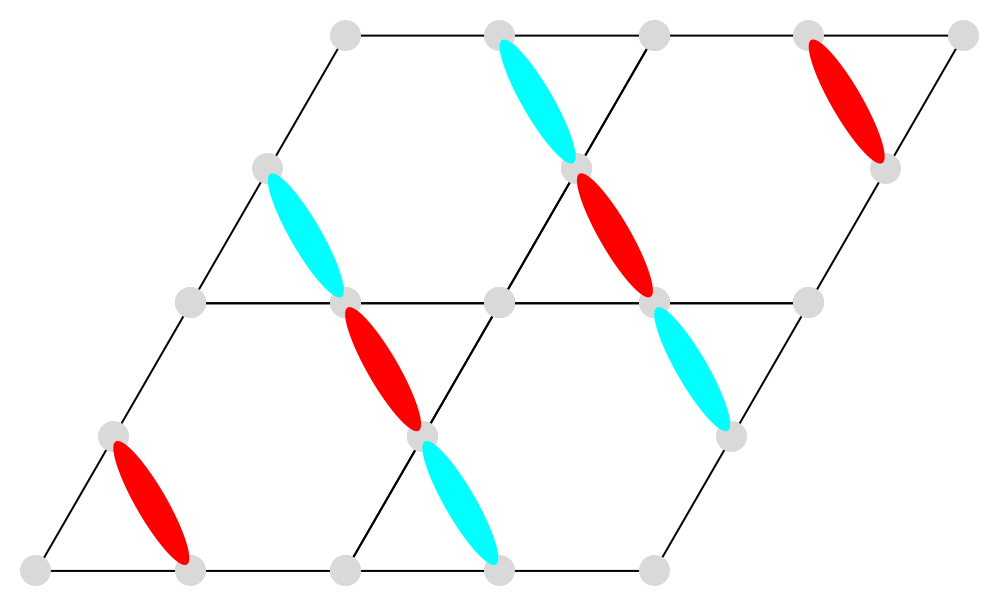} & \includegraphics[width=3cm]{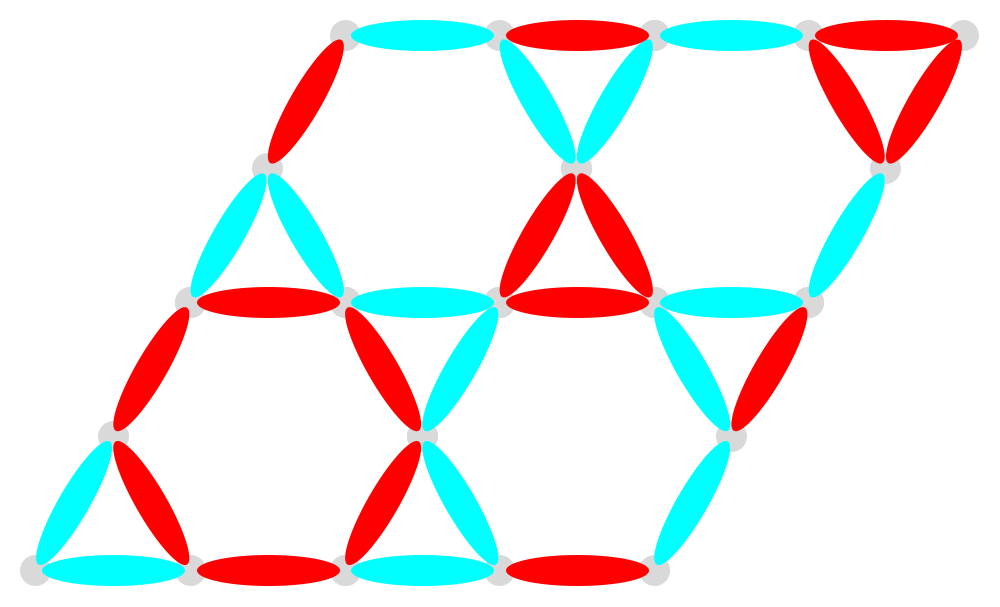}\\
$F_{1}^{(2)}$ & $F_{1,1}^{(2)}$ & $F_{1,2}^{(2)}$ & $F_{1,3}^{(2)}$ &\\
& \includegraphics[width=3cm]{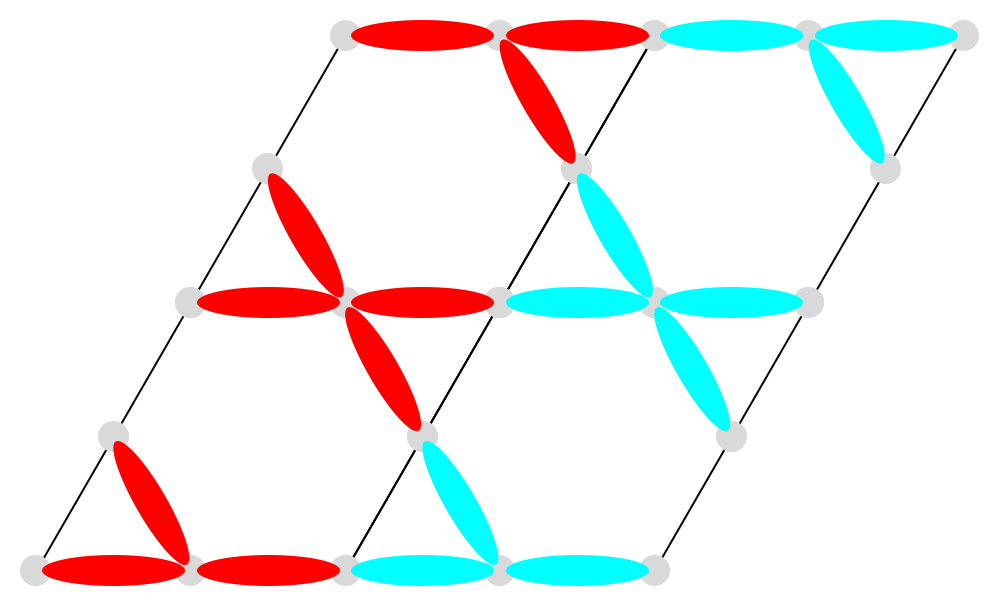} & \includegraphics[width=3cm]{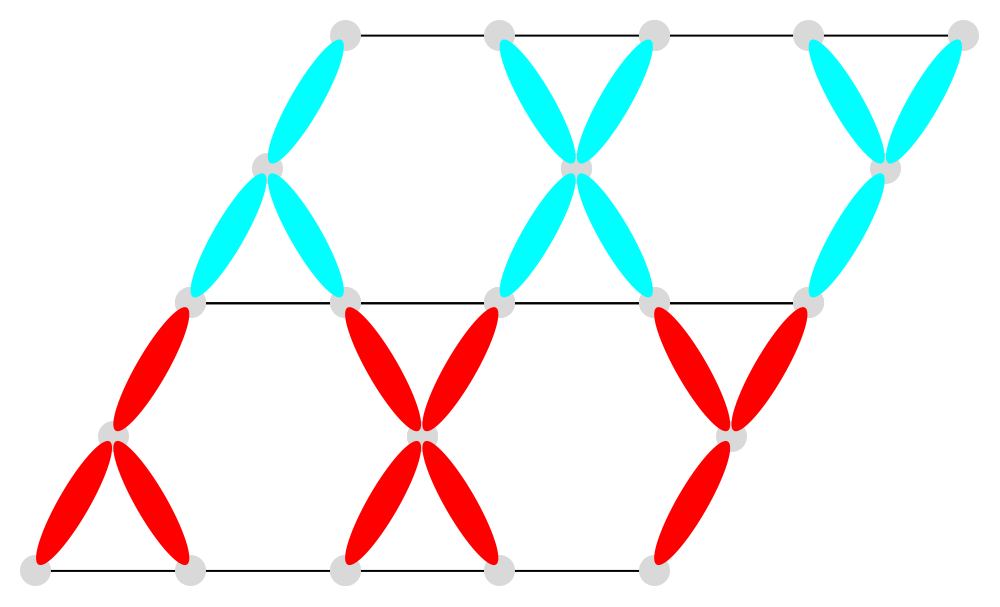} & \includegraphics[width=3cm]{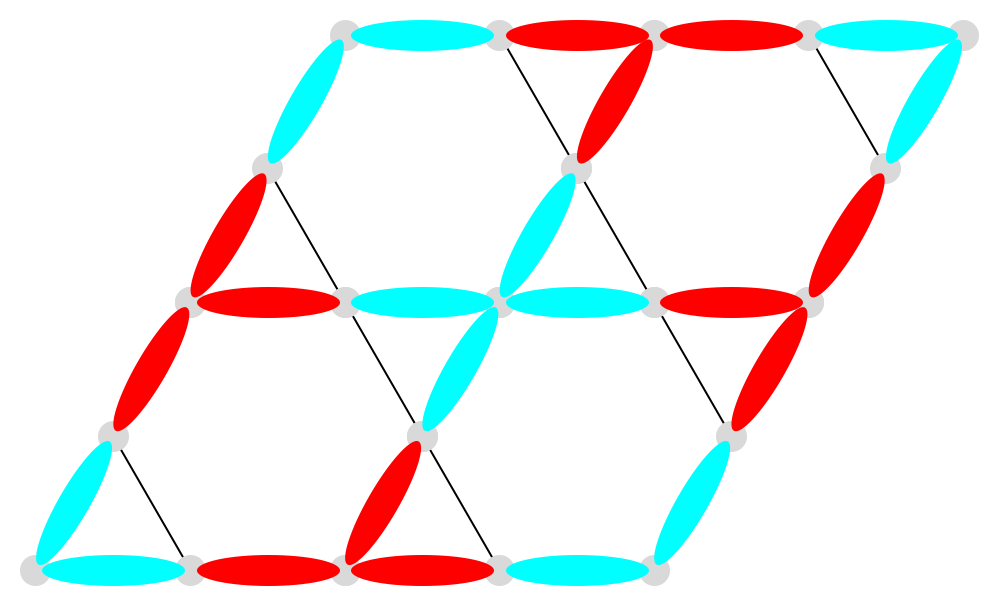} & \includegraphics[width=3cm]{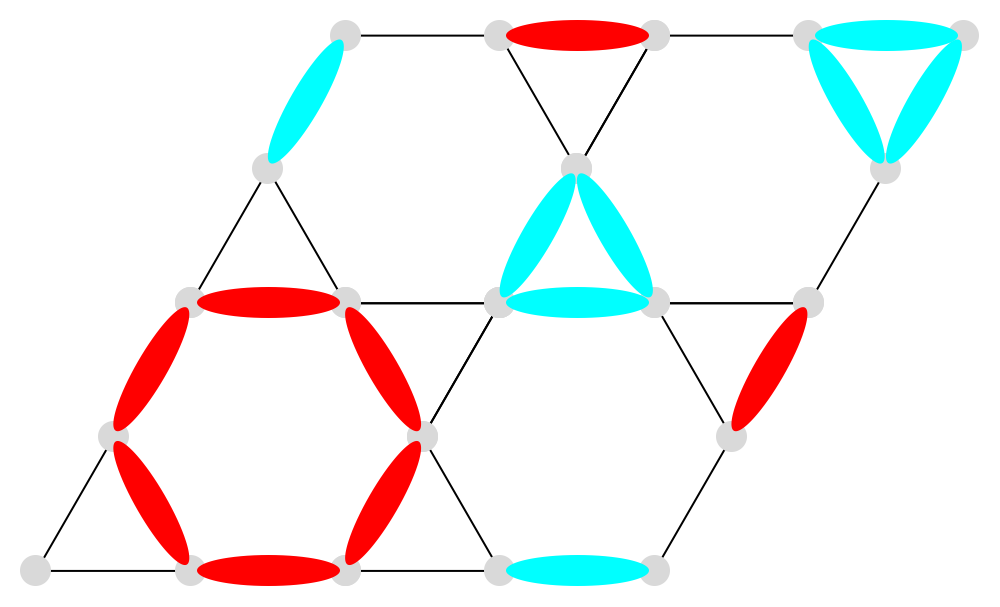}\\
$F_{2}$ & $F_{2,1}$ & $F_{2,2}$ & $F_{2,3}$ &\\
& \includegraphics[width=3cm]{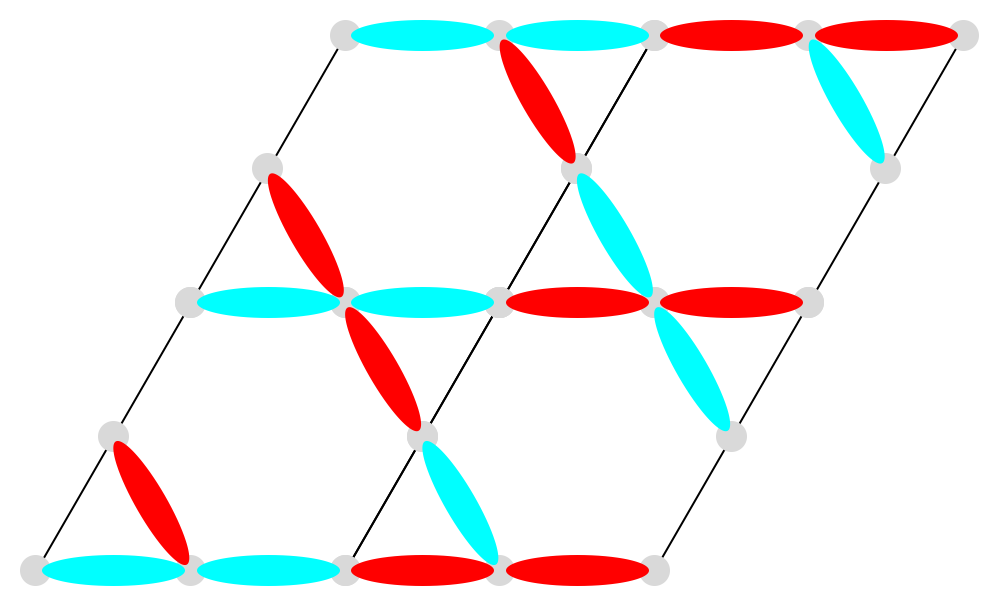} & \includegraphics[width=3cm]{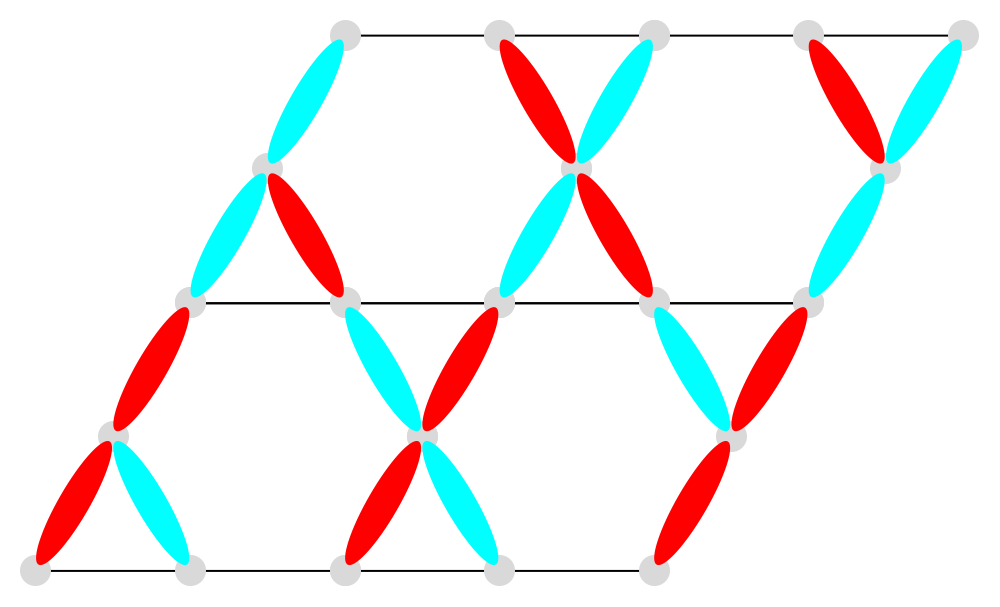} & \includegraphics[width=3cm]{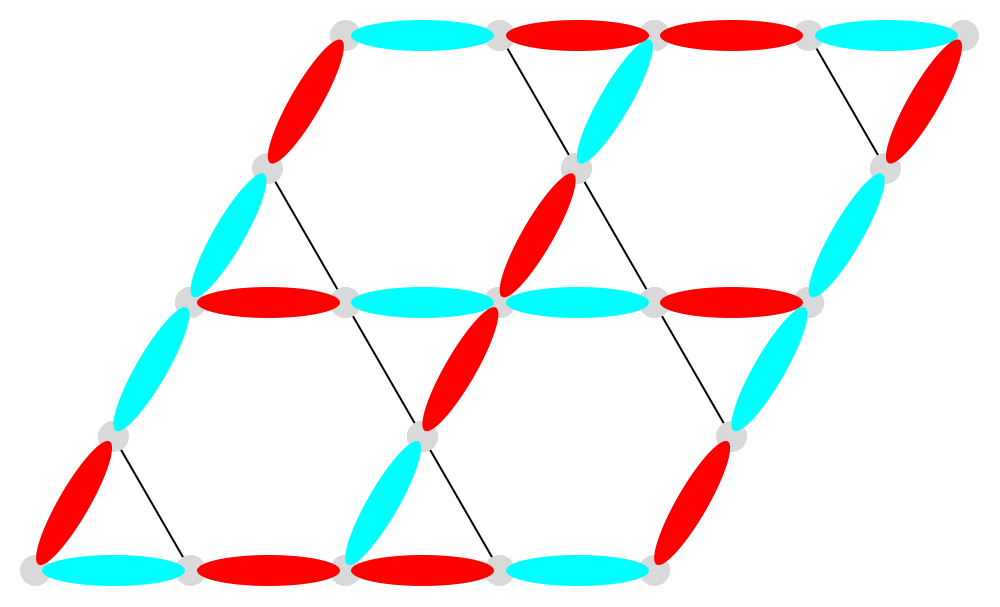} & \includegraphics[width=3cm]{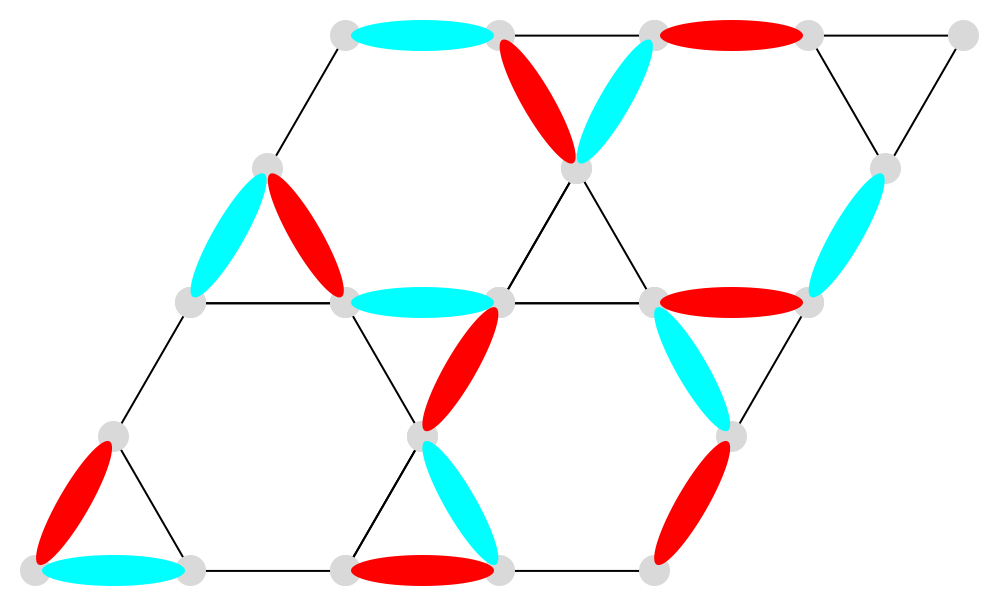}\\
$F_{3}^{(1)}$ & $F_{3,1}^{(1)}$ & $F_{3,2}^{(1)}$ & $F_{3,3}^{(1)}$ &\\
& \includegraphics[width=3cm]{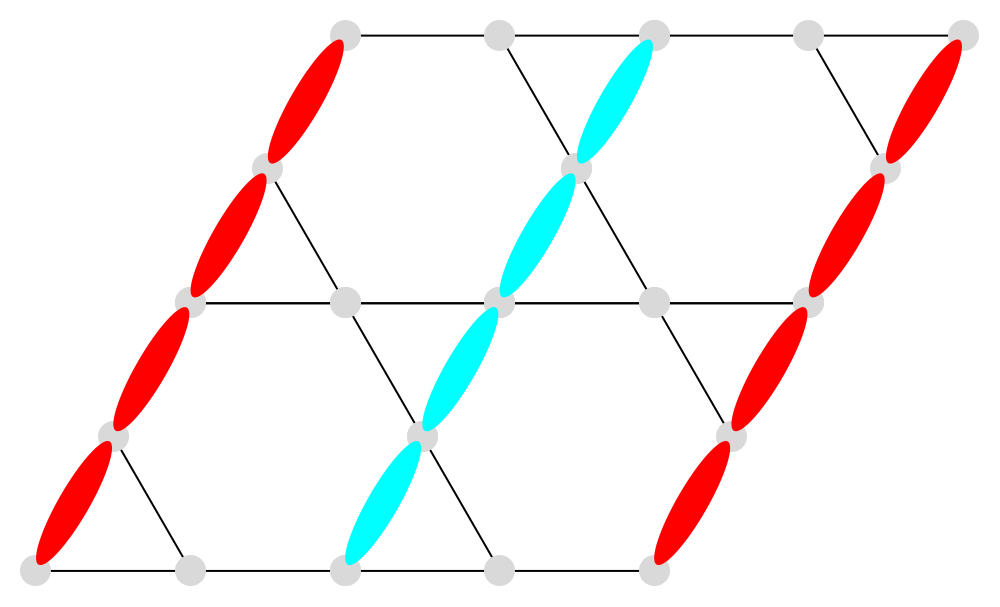} & \includegraphics[width=3cm]{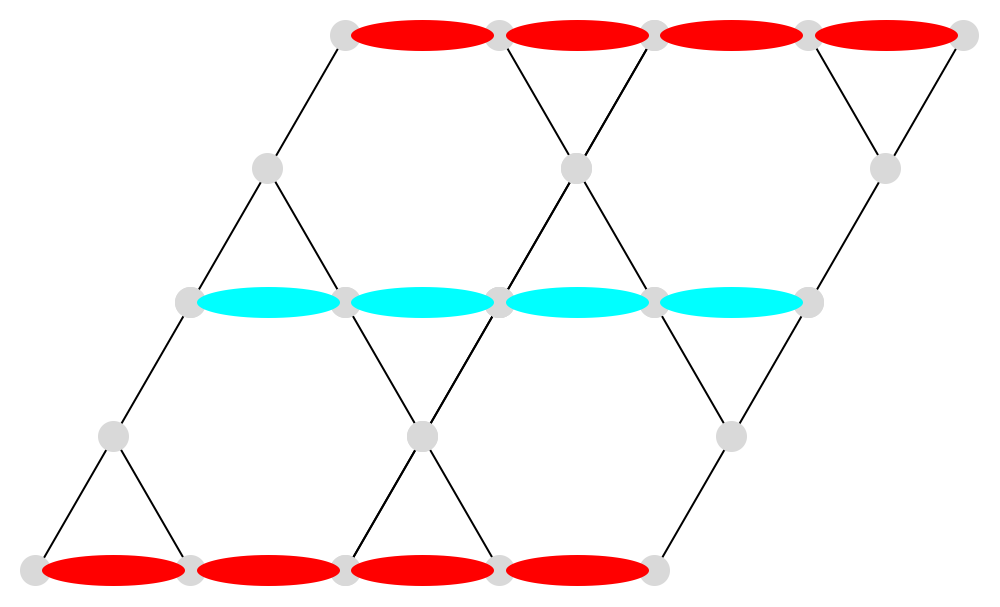} & \includegraphics[width=3cm]{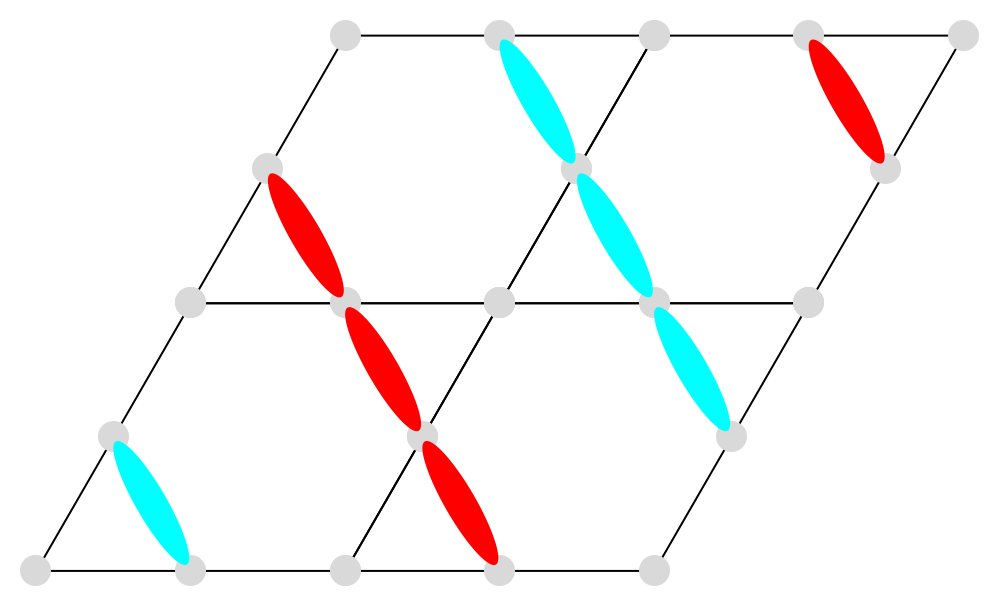} & \includegraphics[width=3cm]{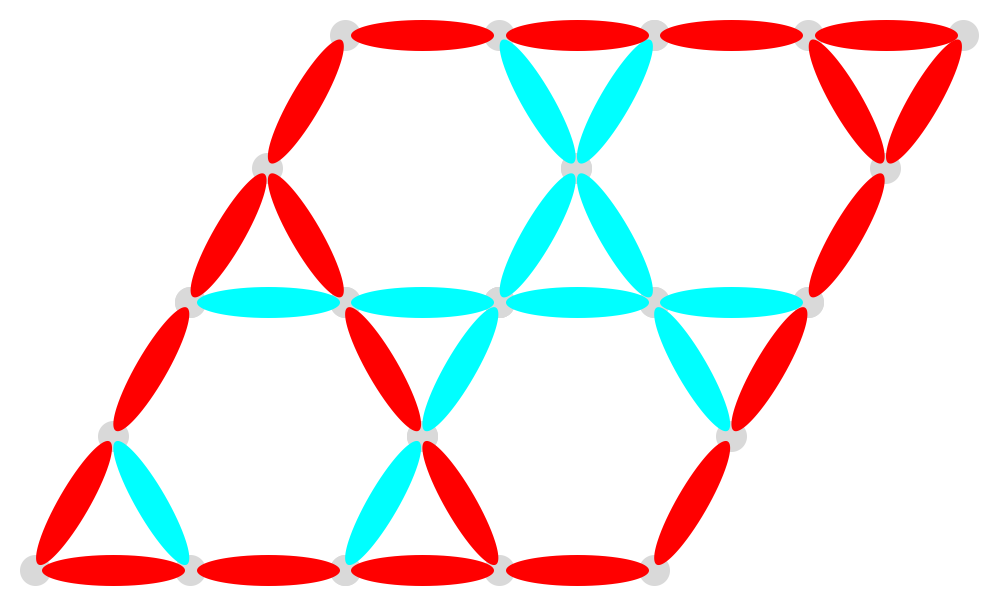}\\
$F_{3}^{(2)}$ & $F_{3,1}^{(2)}$ & $F_{3,2}^{(2)}$ & $F_{3,3}^{(2)}$ &\\
& \includegraphics[width=3cm]{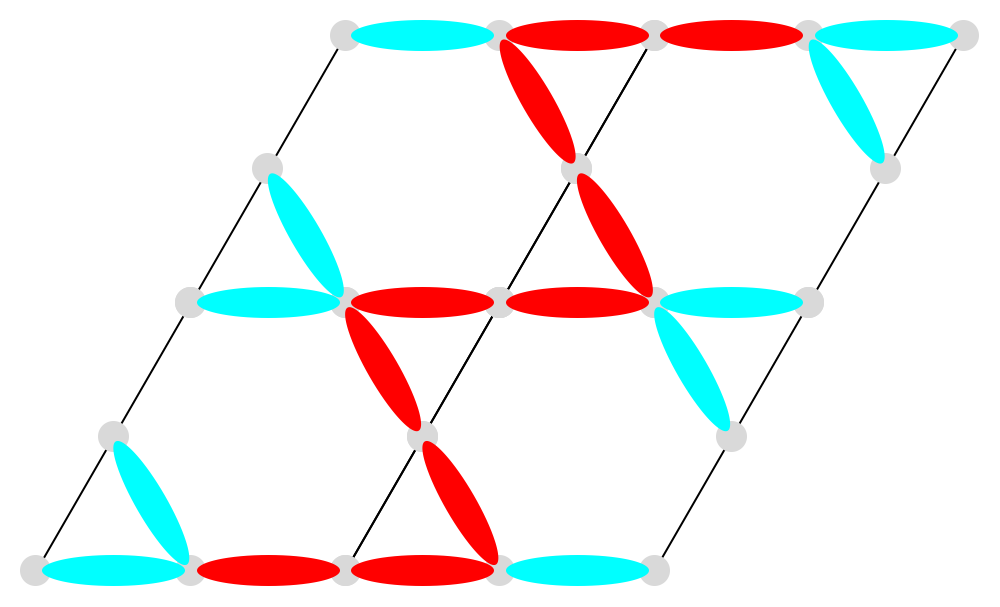} & \includegraphics[width=3cm]{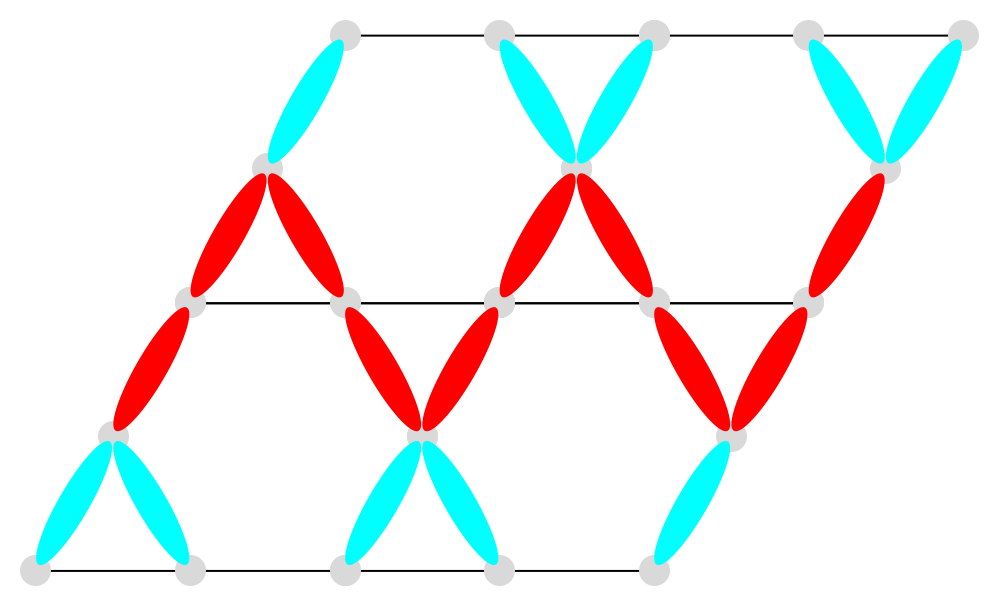} & \includegraphics[width=3cm]{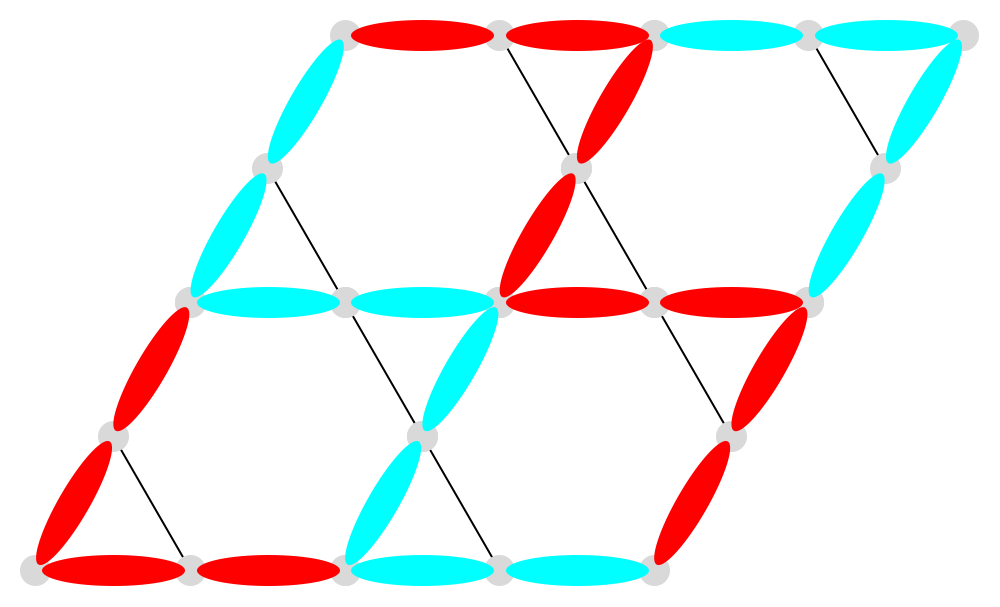} & \includegraphics[width=3cm]{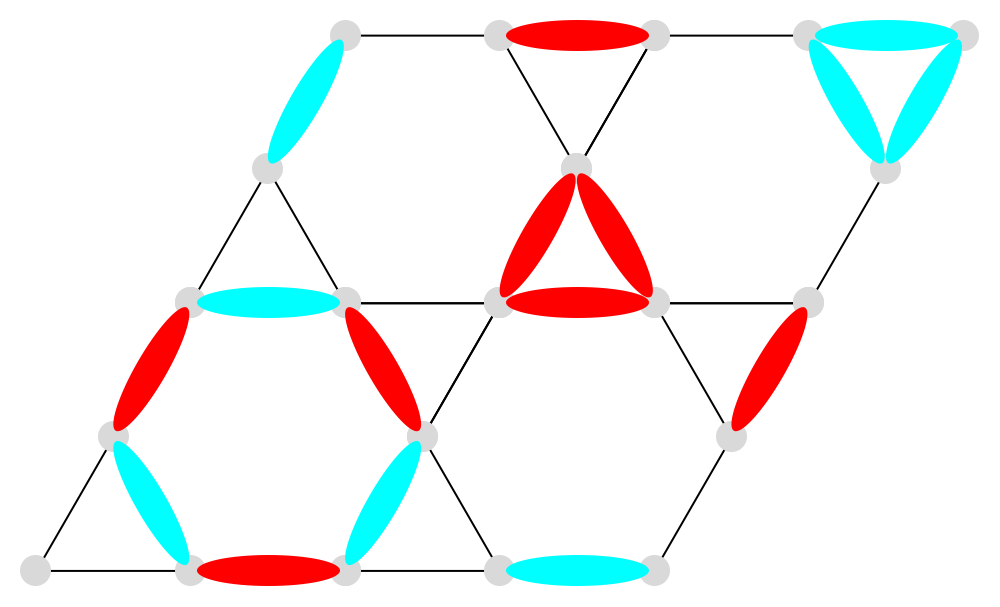}\\
$F_{4}$ & $F_{4,1}$ & $F_{4,2}$ & $F_{4,3}$ &\\
& \includegraphics[width=3cm]{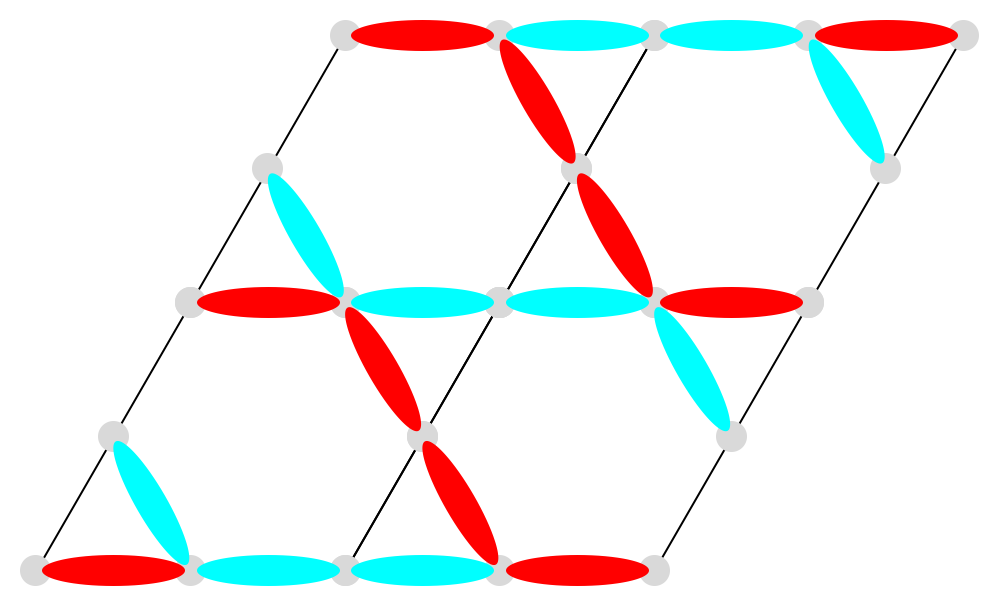} & \includegraphics[width=3cm]{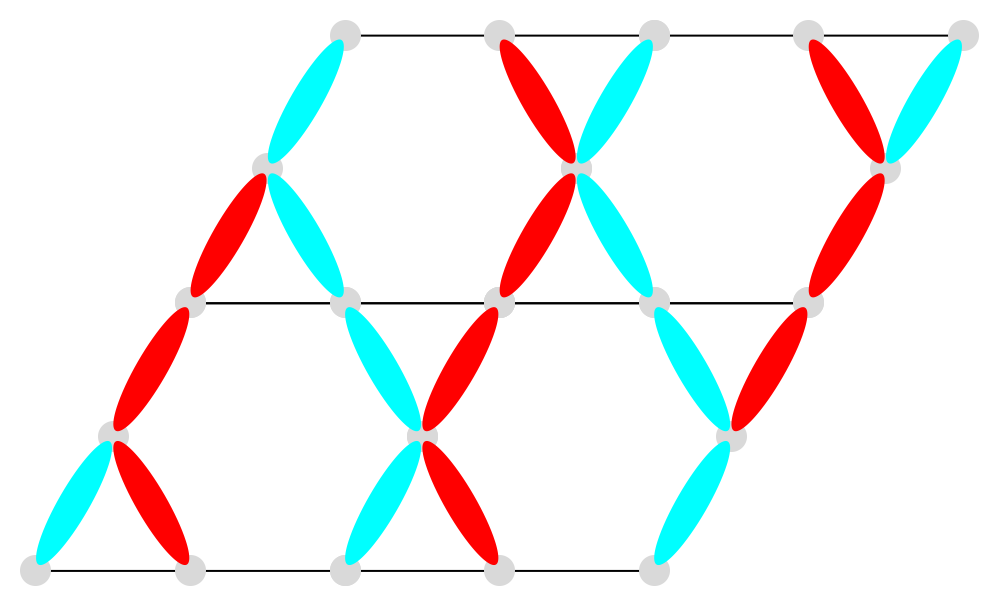} & \includegraphics[width=3cm]{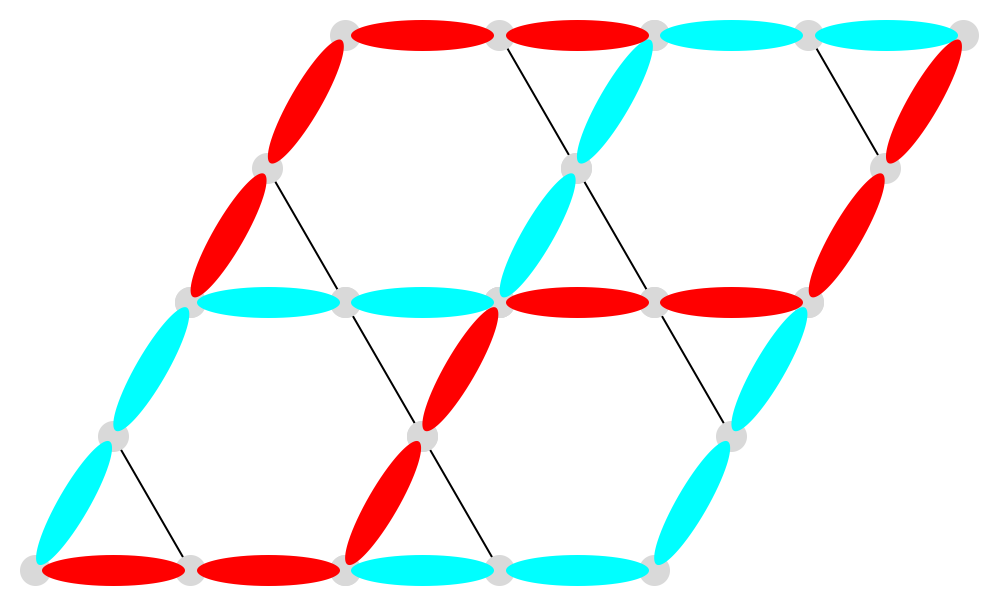} & \includegraphics[width=3cm]{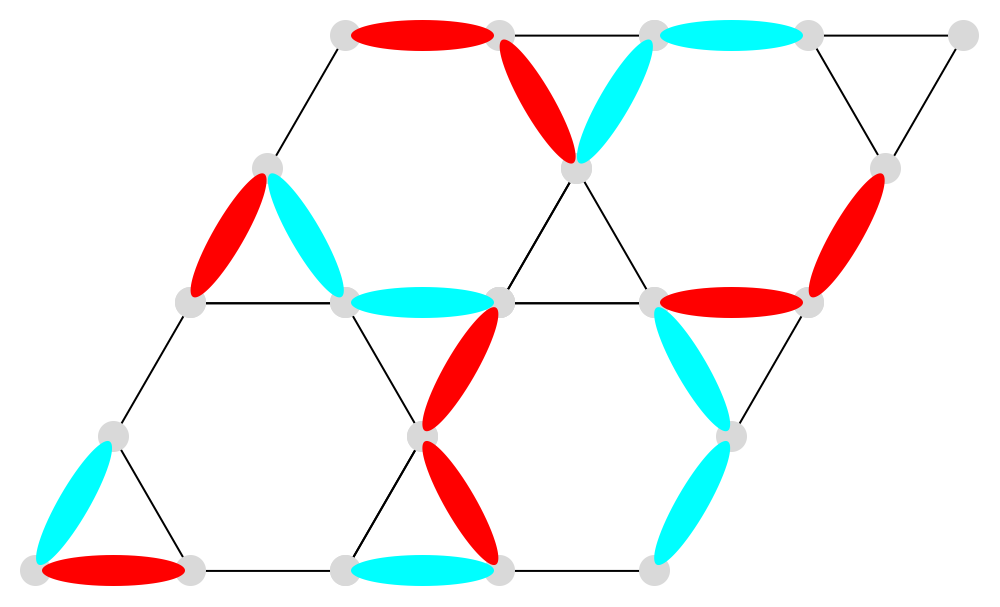}\\
\bottomrule
\end{tabular}}
\caption{\label{tab:NN_SC_OP}Real space illustrations of nearest-neighbor singlet pairing states on the kagome lattice which break translational symmetry. The color (red/blue) denotes the sign ($+1$/$-1$) of the singlet pairing interaction. The first column provides the three-dimensional irrep, denoted according to the $C_{6v}'''$ point group. The second column shows each component of the irrep, and the last column shows a symmetric superposition of the three components.}
\end{center}
\end{table}

\begin{table}[h!]
\begin{center}
\resizebox{0.9\columnwidth}{!}{
\begin{tabular}{ccccc}
\toprule
\textbf{Irrep} & \multicolumn{3}{c}{\textbf{Components}} & \textbf{Symmetric superposition}\\
\hline
$F_{1}$ & $F_{1,1}$ & $F_{1,2}$ & $F_{1,3}$ &\\
& \includegraphics[width=3cm]{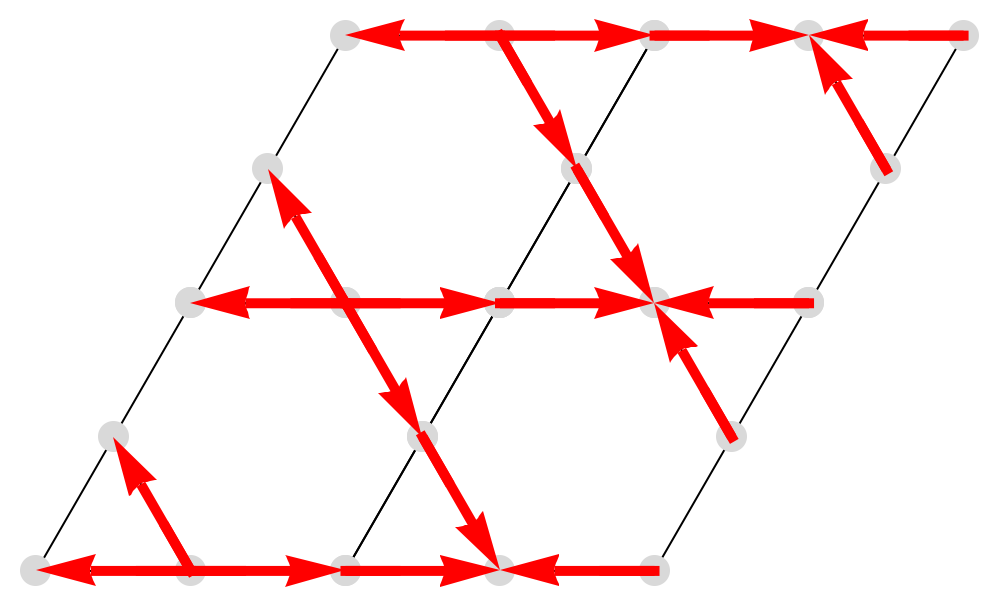} & \includegraphics[width=3cm]{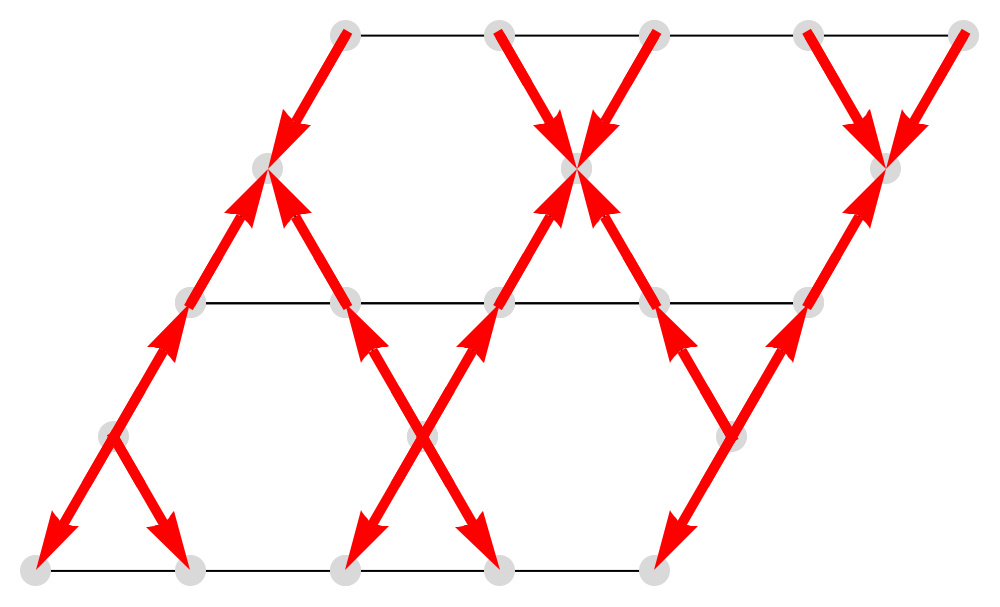} & \includegraphics[width=3cm]{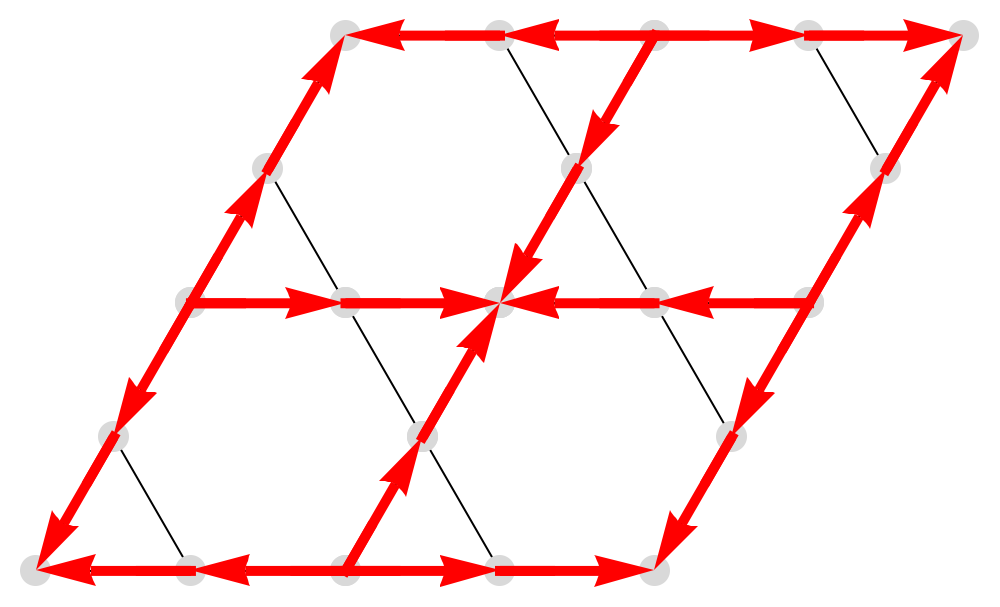} & \includegraphics[width=3cm]{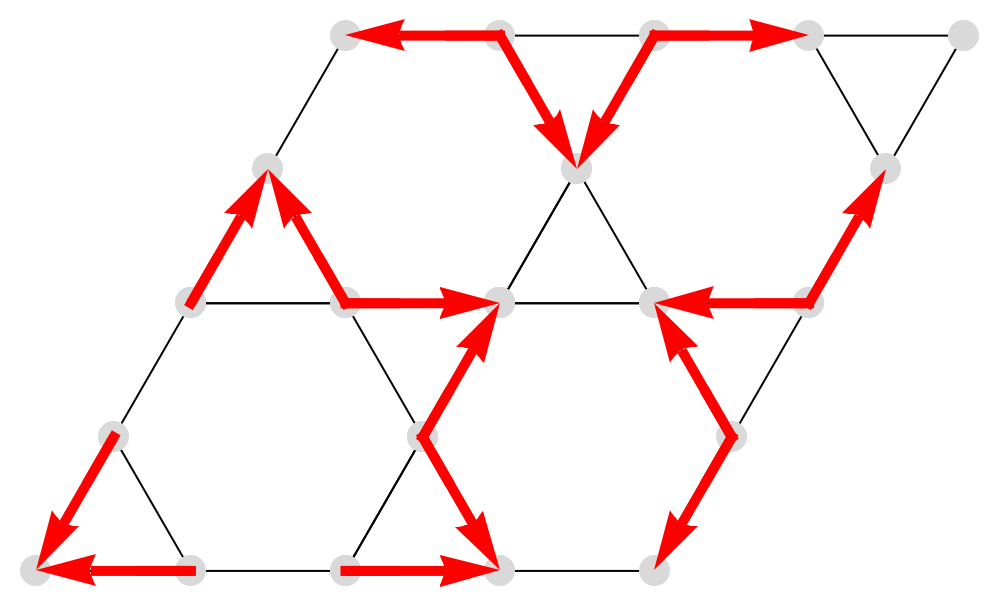}\\
$F_{2}^{(1)}$ & $F_{2,1}^{(1)}$ & $F_{2,2}^{(1)}$ & $F_{2,3}^{(1)}$ &\\
& \includegraphics[width=3cm]{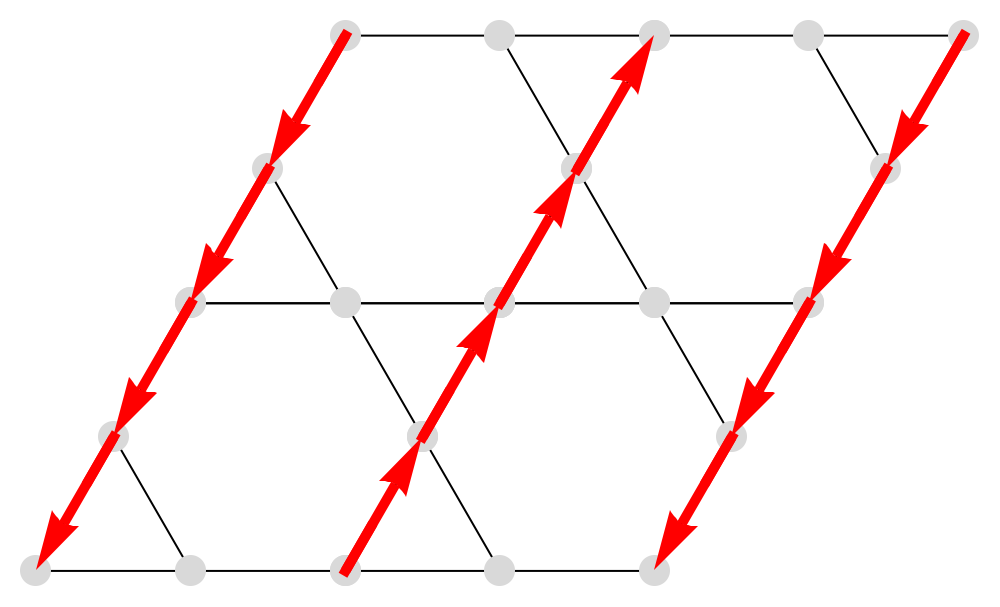} & \includegraphics[width=3cm]{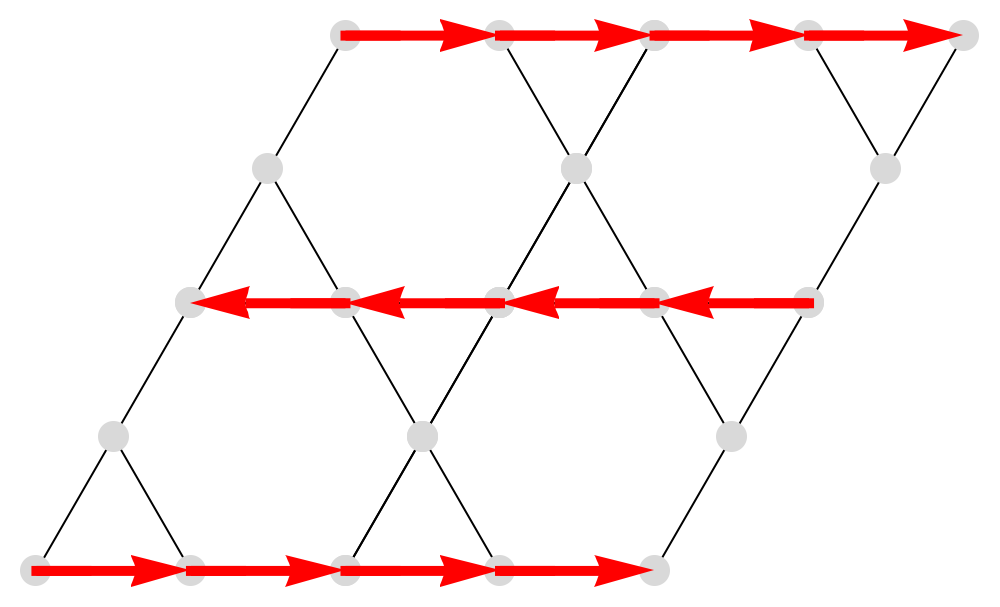} & \includegraphics[width=3cm]{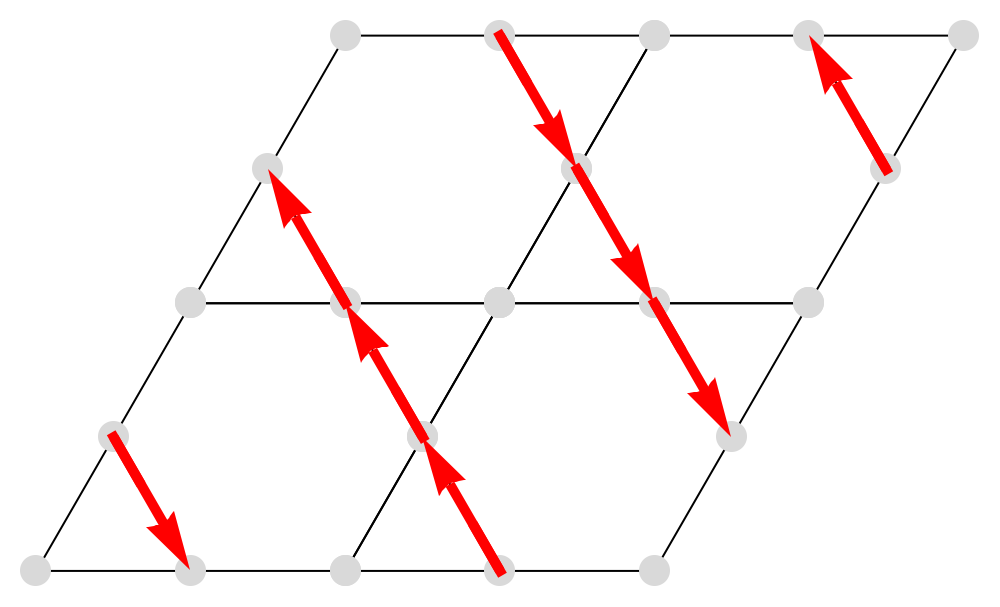} & \includegraphics[width=3cm]{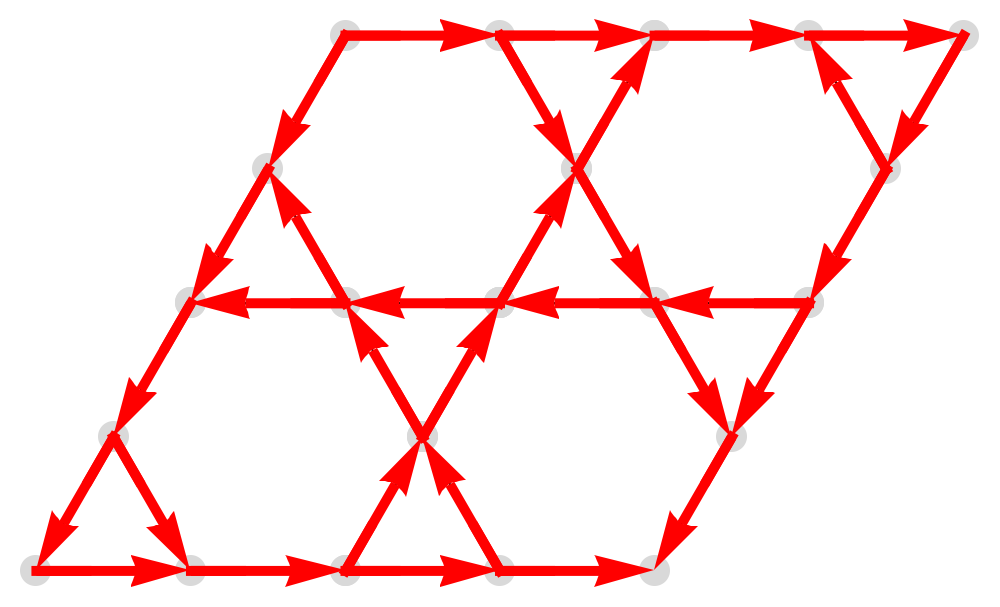}\\
$F_{2}^{(2)}$ & $F_{2,1}^{(2)}$ & $F_{2,2}^{(2)}$ & $F_{2,3}^{(2)}$ &\\
& \includegraphics[width=3cm]{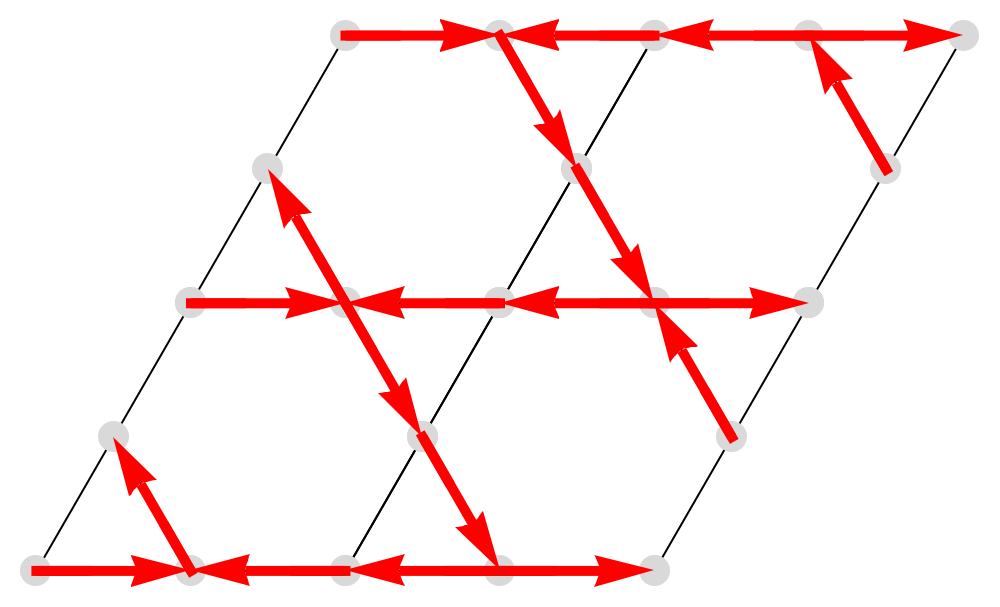} & \includegraphics[width=3cm]{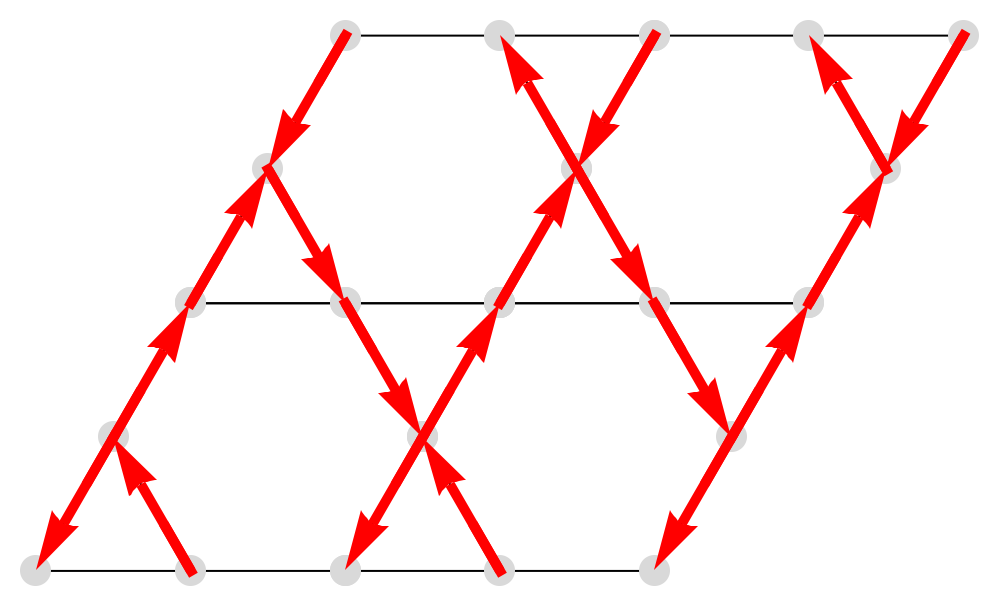} & \includegraphics[width=3cm]{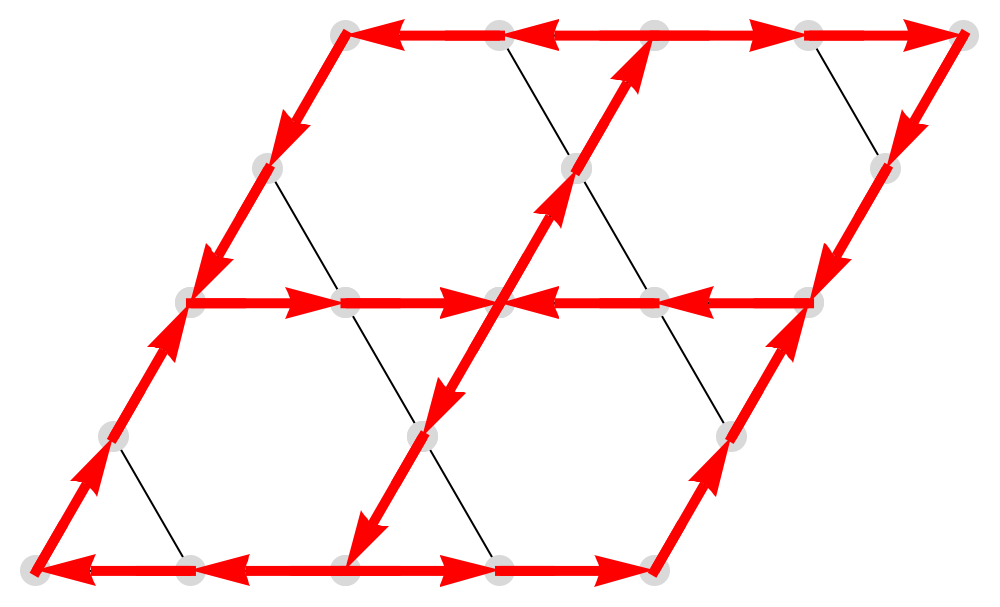} & \includegraphics[width=3cm]{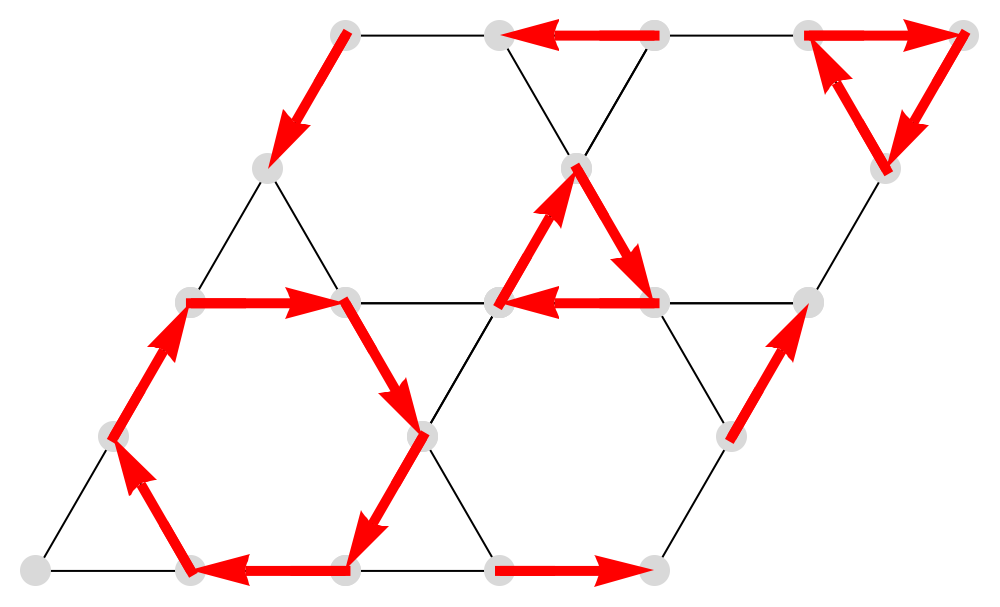}\\
$F_{3}$ & $F_{3,1}$ & $F_{3,2}$ & $F_{3,3}$ &\\
& \includegraphics[width=3cm]{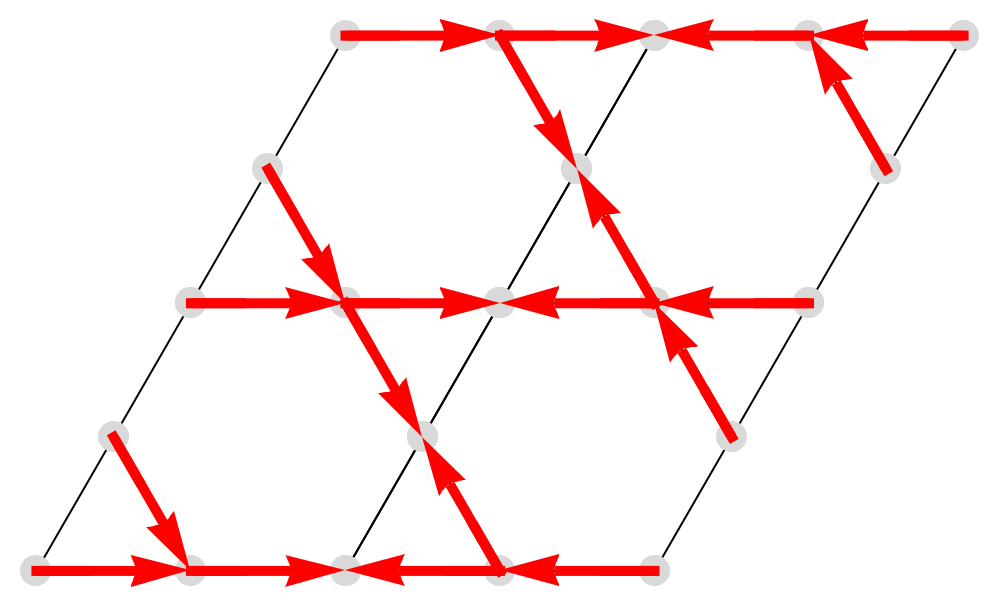} & \includegraphics[width=3cm]{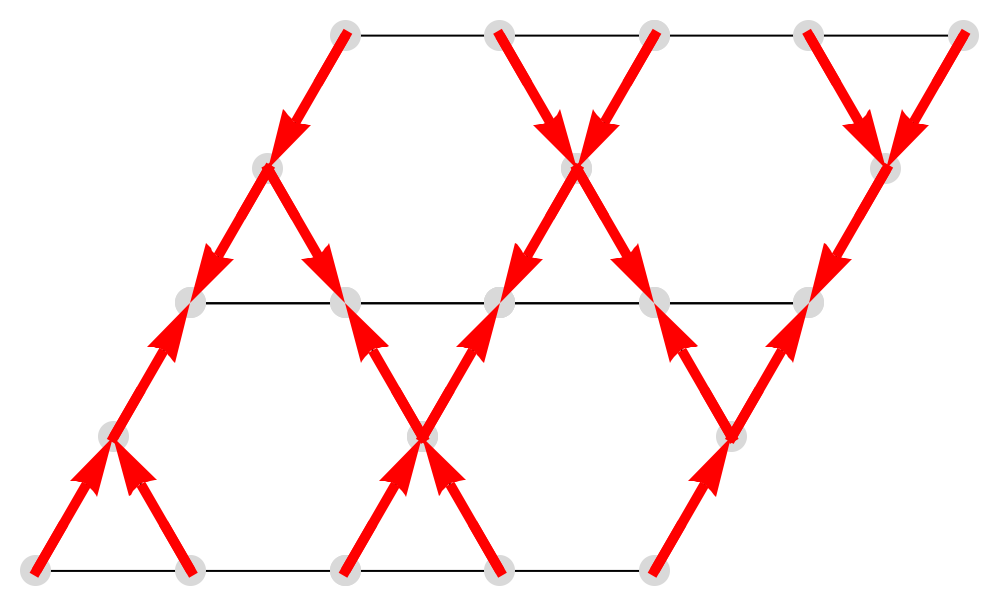} & \includegraphics[width=3cm]{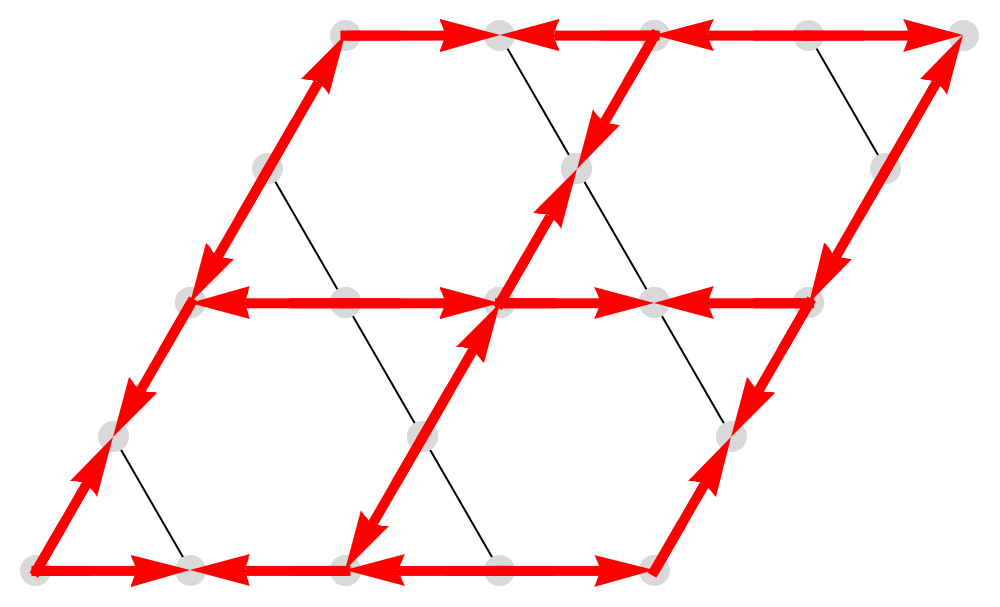} & \includegraphics[width=3cm]{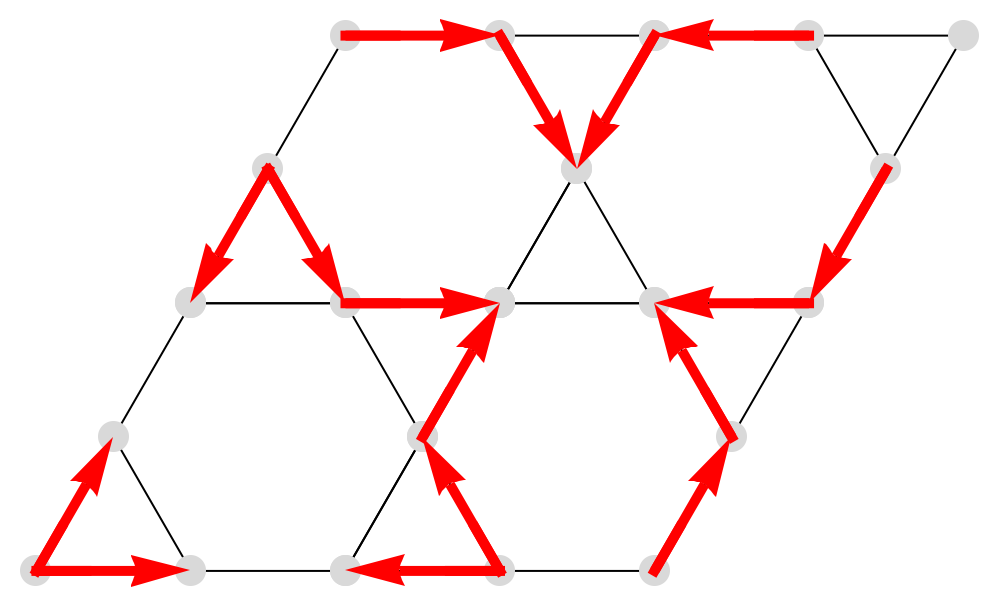}\\
$F_{4}^{(1)}$ & $F_{4,1}^{(1)}$ & $F_{4,2}^{(1)}$ & $F_{4,3}^{(1)}$ &\\
& \includegraphics[width=3cm]{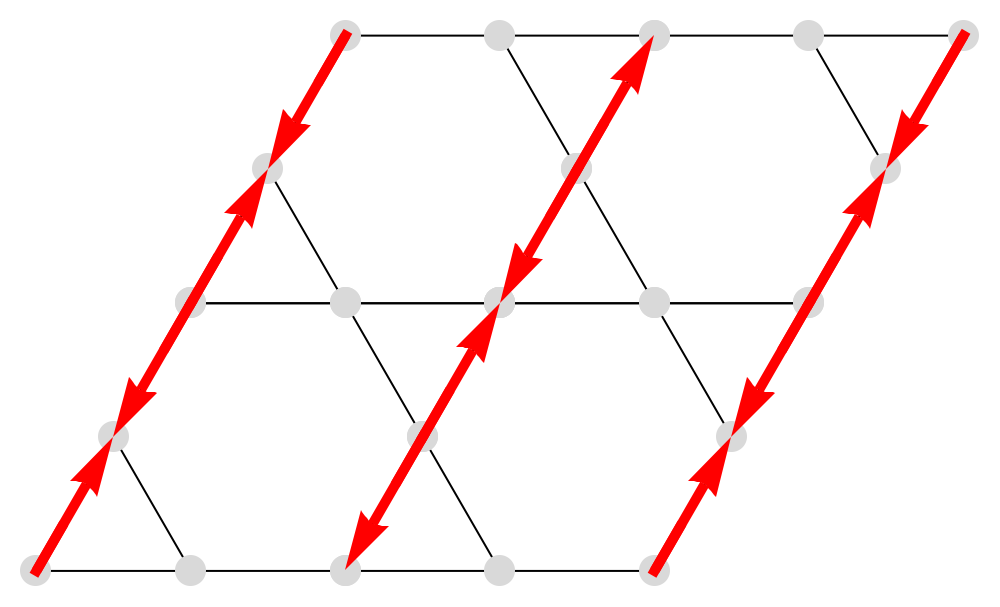} & \includegraphics[width=3cm]{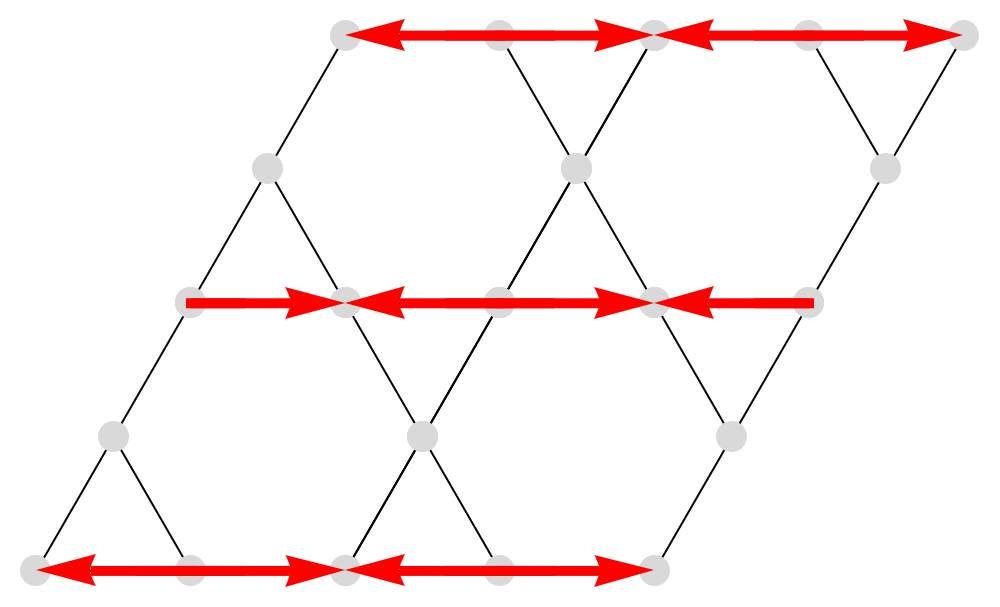} & \includegraphics[width=3cm]{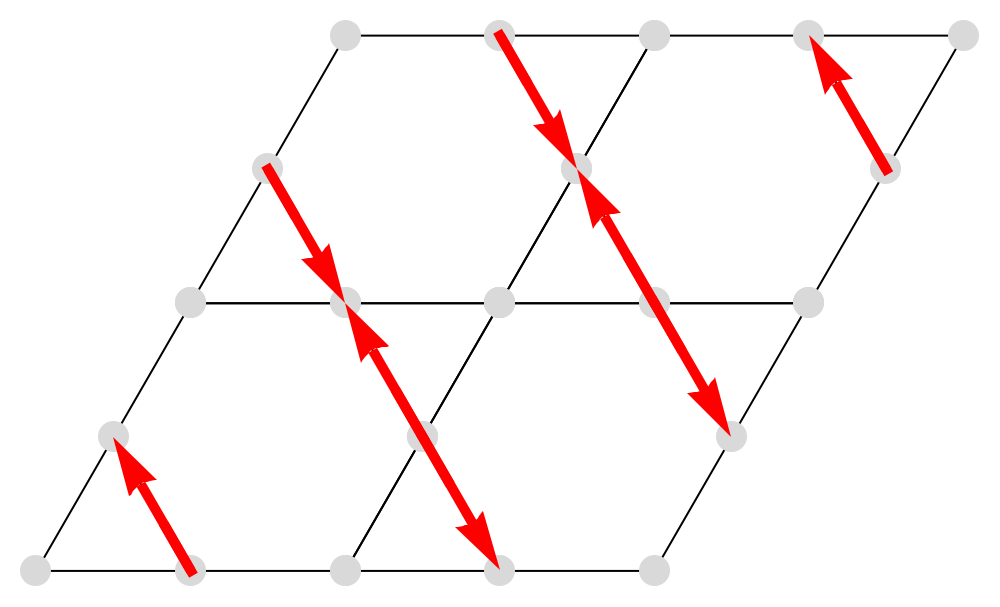} & \includegraphics[width=3cm]{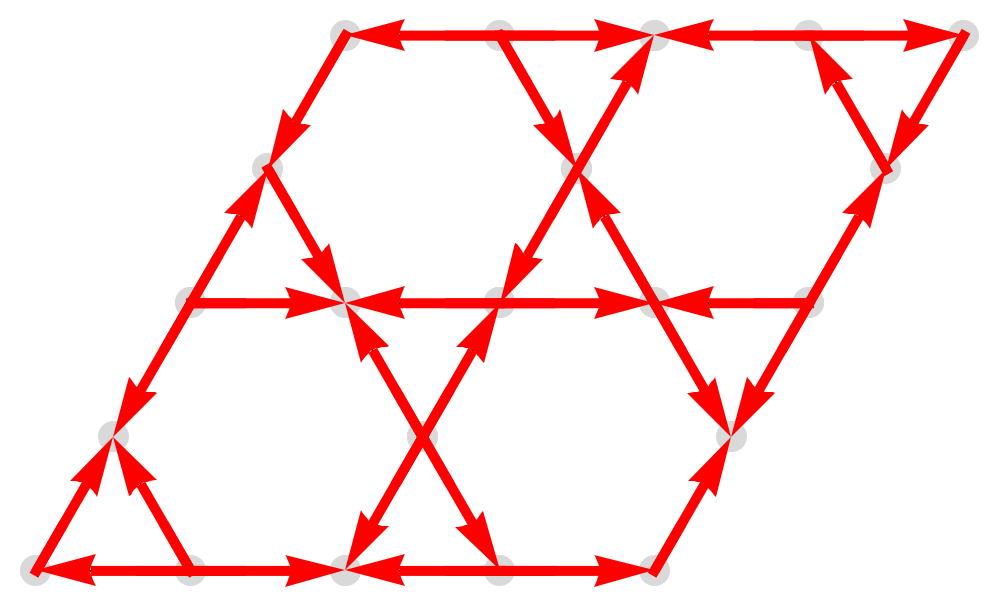}\\
$F_{4}^{(2)}$ & $F_{4,1}^{(2)}$ & $F_{4,2}^{(2)}$ & $F_{4,3}^{(2)}$ &\\
& \includegraphics[width=3cm]{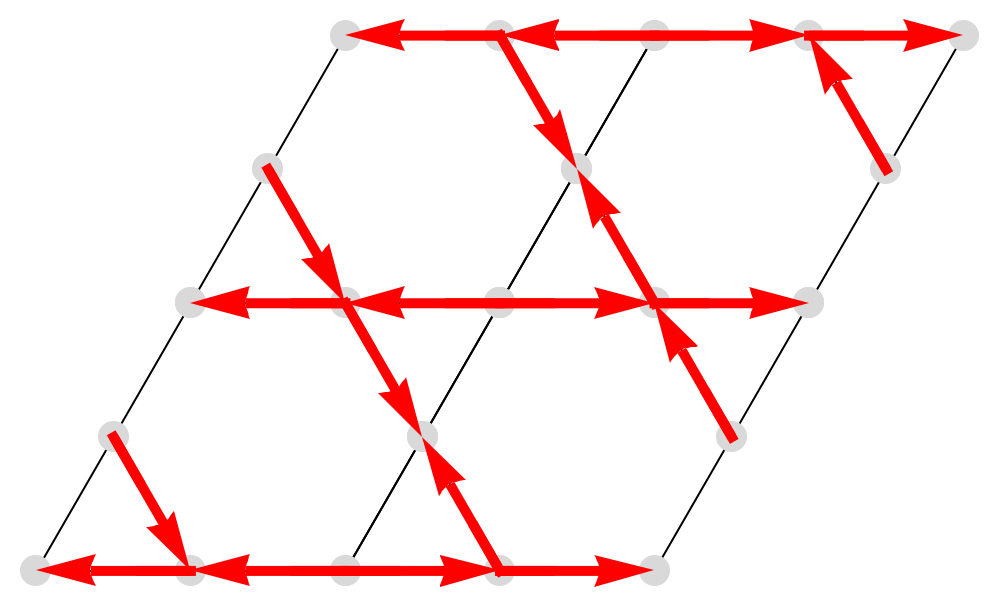} & \includegraphics[width=3cm]{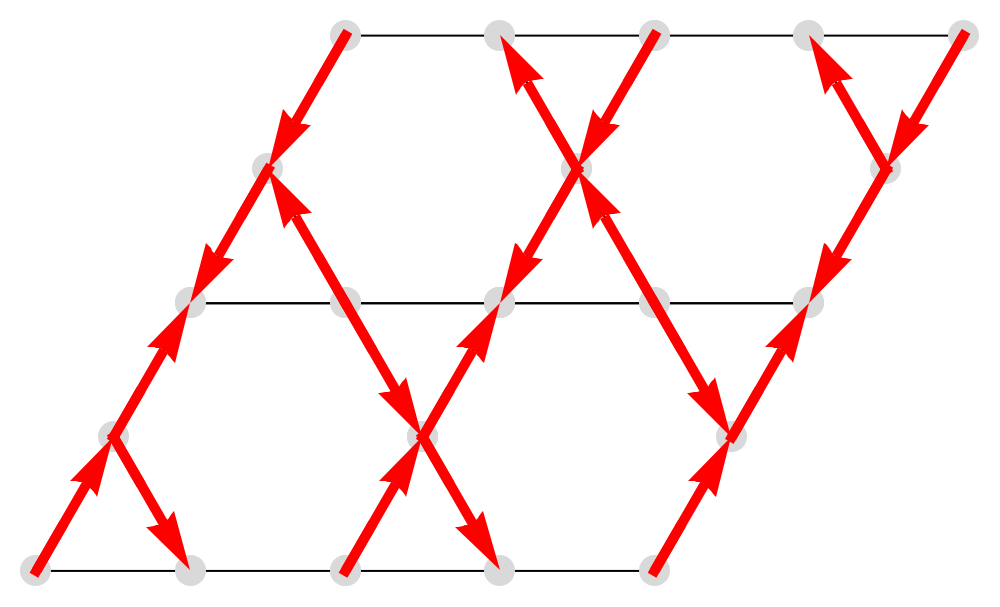} & \includegraphics[width=3cm]{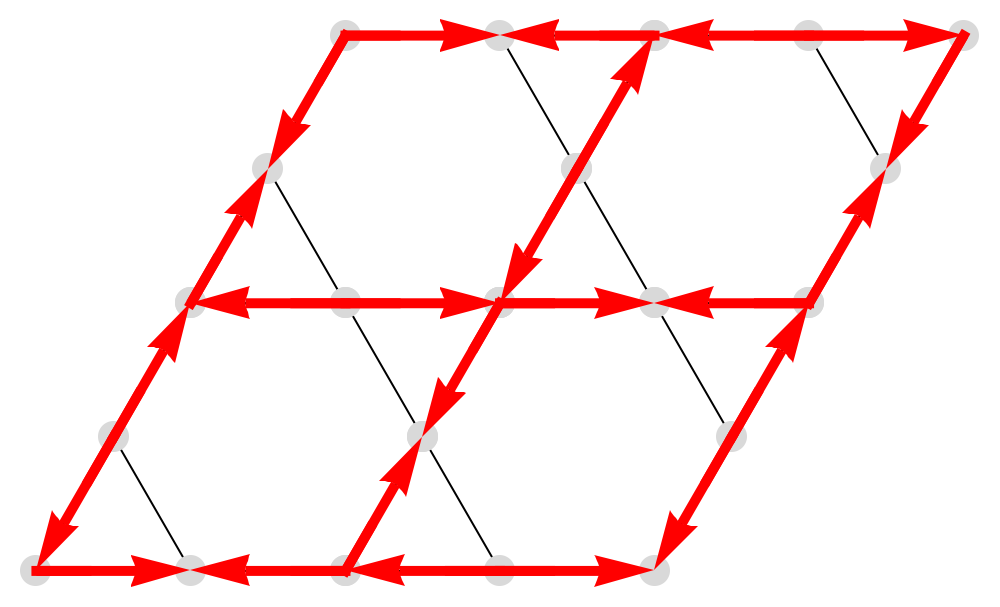} & \includegraphics[width=3cm]{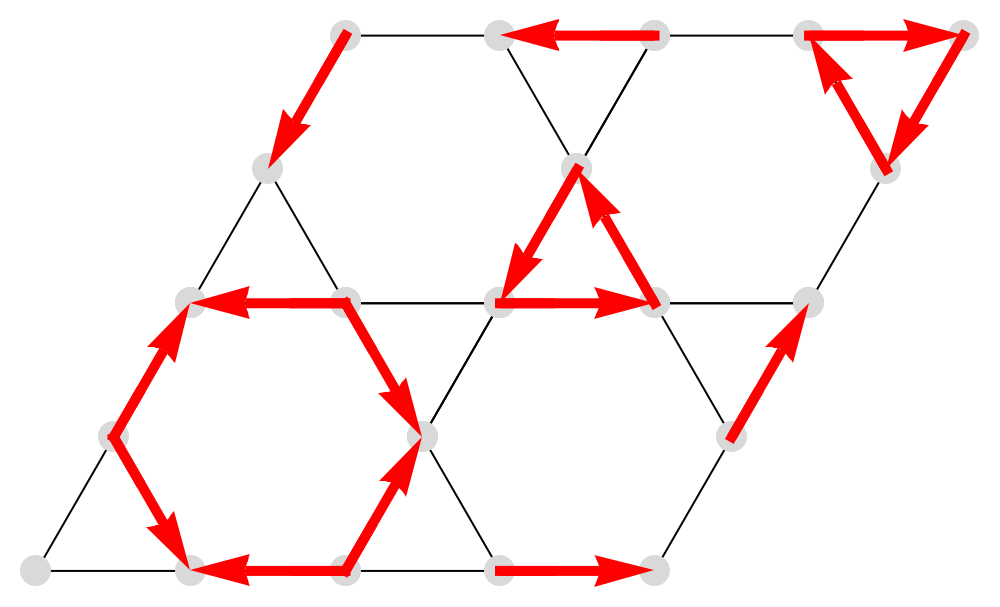}\\
\bottomrule
\end{tabular}}
\caption{\label{tab:NN_triplet_SC_OP}Real space illustrations of nearest-neighbor triplet pairing states on the kagome lattice which break translational symmetry. The first column provides the three-dimensional irrep, denoted according to the $C_{6v}'''$ point group. The second column shows each component of the irrep, and the last column shows a symmetric superposition of the three components.}
\end{center}
\end{table}
\end{widetext}

\bibliography{Kagome}
\end{document}